\documentclass[10pt, twocolumn]{article}

\usepackage{booktabs}
\usepackage[switch]{lineno} 
\usepackage{graphicx}
\usepackage{amsmath}
\usepackage{amssymb}
\usepackage{geometry}
\usepackage[labelfont=bf]{caption}
\usepackage{amssymb,amsmath,graphicx,color,amsfonts}
\usepackage{bm}
\usepackage[tight]{subfigure}
\usepackage[export]{adjustbox}
\usepackage{braket}
\usepackage[colorlinks=true,citecolor=blue,linkcolor=blue]{hyperref}
\usepackage{cancel}
\usepackage{multirow}
\usepackage{blindtext}

\newcounter{oldtocdepth}

\usepackage[normalem]{ulem}
\usepackage{xcolor}

\newcommand\beq{\begin{equation}}
\newcommand\eeq{\end{equation}}
\newcommand\bea{\begin{eqnarray}}
\newcommand\eea{\end{eqnarray}}

\renewcommand{\figurename}{\textbf{Fig.}}

\title{\bfseries 
Observation of Robust and Coherent Non-Abelian Hadron Dynamics on Noisy Quantum Processors
}

\author{Fran Ilčić$^{1,*}$, Ritajit Majumdar$^{2,*}$,  Emil Mathew$^{1,*}$, Md. Osama Ali$^{1}$,\\ Nathan Earnest-Noble$^{3}$ \& Indrakshi Raychowdhury$^{1,**}$}

\date{}
\begin{document}

\maketitle
\vspace{1.5cm}

\noindent\small{$^1$Department of Physics and Center for Research in Quantum Information and Technology,\\ Birla Institute of Technology and Science Pilani,\\ K K Birla Goa Campus, Zuarinagar, Sancole, Goa  403726, India. \\
$^2$IBM Quantum, IBM India Research Lab, India.\\
$^3$IBM Quantum, IBM T.J. Watson Research Center, Yorktown Heights, NY 10598, USA.\\
$^*$ contributed equally.\\
$**$ corresponding author. email: indrakshir@goa.bits-pilani.ac.in


\section*{~~~~}
\textbf{The real-time evolution of strongly interacting matter remains a frontier of fundamental physics, as classical simulations are hampered by exponential Hilbert space growth and rapid, unmanageable growth of quantum entanglement.  This study reports the quantum simulation of hadron dynamics within a $(1+1)$-dimensional SU(2) lattice gauge theory using a 156-qubit IBM superconducting processor.
Leveraging a hardware-efficient Loop-String-Hadron (LSH) encoding, we simulate the dynamics of the physical degrees of freedom on a $60$-site lattice in the weak-coupling regime, as a crucial step toward the continuum limit. The hardware data reveal confined meson propagation and early-time oscillations of the mesonic profile, from which we extract a breathing-mode frequency as a spectroscopic observable. Benchmarking against tensor-network simulations of the full LSH Hamiltonian and Pauli-propagation simulations of the noiseless circuit supports the validity of the physical approximation, the quantum algorithm and the observed dynamics within the accessible time window. These results show that physics-native encodings can enable scalable access to coherent non-Abelian real-time dynamics on noisy quantum hardware.}

\section*{~~~~}

Gauge theories form the bedrock of the Standard Model, yet predicting their non-equilibrium dynamics remains notoriously difficult. While Euclidean Monte Carlo methods successfully compute static properties, they cannot tackle real-time evolution due to the sign problem. Tensor Network (TN) methods offer a partial solution in low dimensions but hit a fundamental ``entanglement wall'' during quenches: as time evolves, the entanglement entropy $S(t)$ grows linearly, requiring the bond dimension to grow exponentially to maintain accuracy. This renders long-time simulations of large lattices computationally prohibitive for classical machines.


Quantum simulation, utilizing either digital gate-based or analog approaches, provides a fundamental route around this barrier by mapping gauge fields directly onto controllable quantum degrees of freedom. However, scalable experimental demonstrations have been predominantly confined to Abelian models, such as U(1) or $\mathbb{Z}_2$ lattice gauge theories \cite{Martinez:2016yna, Yang:2020yer, Mil:2019pbt, Mildenberger:2022jqr, Meth:2023wzd, Cochran:2024rwe, Zhou:2021kdl, Mueller:2024mmk, Schweizer:2019lwx, Mueller:2022xbg, Farrell:2024fit, Davoudi:2024wyv, Cobos:2025krn, Gonzalez-Cuadra:2024xul, Xiang:2025qhq, Schuhmacher:2025ehh}. The transition to non-Abelian symmetries \cite{Klco:2019evd, Ciavarella:2021nmj, Atas:2021ext, Atas:2022dqm, Ciavarella:2024fzw,Farrell:2022wyt, Farrell:2022vyh, Chernyshev:2025lil}, a strictly necessary step for simulating SU(3) and the phenomenology of Quantum Chromodynamics (QCD) \cite{Bauer:2023qgm, DiMeglio:2023nsa}, remains stalled by the difficulty of enforcing non-commuting Gauss's laws on physical hardware. While analog simulators often lack the flexibility to implement these complex local constraints, current digital quantum devices face a parallel challenge: the deep circuits required to enforce gauge invariance are typically overwhelmed by hardware noise. Accessing the dynamics over longer timescales and reaching the continuum, or even the thermodynamic limit, remained a long-term goal for the community. 

In this work, we address this challenge using the Loop–String–Hadron (LSH) framework  \cite{Raychowdhury:2019iki}, which reorganizes the local Hilbert space directly in terms of gauge-invariant degrees of freedom and avoids the non-local structures that complicate standard encodings. With this hardware-efficient encoding approach, we are able to scale the simulation of the dynamics of a meson in $(1+1)$-dimensional lattice gauge theory to a $ 60$-staggered-site SU(2) model using a 156-qubit superconducting processor—approaching the thermodynamic limit. The coupling constants are chosen to be in the weak-coupling regime, where entanglement is expected to grow rapidly, while the quantum algorithm involves a constant circuit depth per time step. We simulate dynamics of the system with a noise-resilient measurement protocol \cite{Berg:2020ibi}. A differential measurement is employed to isolate the coherent mesonic signal from the background vacuum dynamics, allowing physically meaningful observables to be extracted directly from noisy hardware data.

To assess the experiment, we compare the hardware data against two complementary classical baselines. First, tensor-network simulations of the full LSH Hamiltonian \cite{Gupta:2026tcg} provide a reference for the physical dynamics over the accessible time window. Second, Pauli-propagation simulations of the noiseless circuit \cite{Rudolph:2025gyq} validate the digital implementation of the approximated dynamics. This layered comparison allows us to separately evaluate the physical approximation, the circuit-level algorithm and the experimental realization. Within the explored regime, we observe confined meson propagation, early-time internal oscillations of the mesonic profile, and a spectroscopic frequency that can be consistently extracted across methods. Compute time of real time dynamics up to each Time step exhibits different scaling properties for different classical and quantum methods.As the weak-coupling regime is approached, the practical cost and control of the available classical baselines become increasingly challenging within fixed resource budgets, while the QPU runtime remains set primarily by the shot budget. These comparisons motivate the detailed analysis presented below.


\section*{The Physics and the Experiment} 
\begin{figure*}[t!]
    \centering
    \includegraphics[width=1\linewidth]{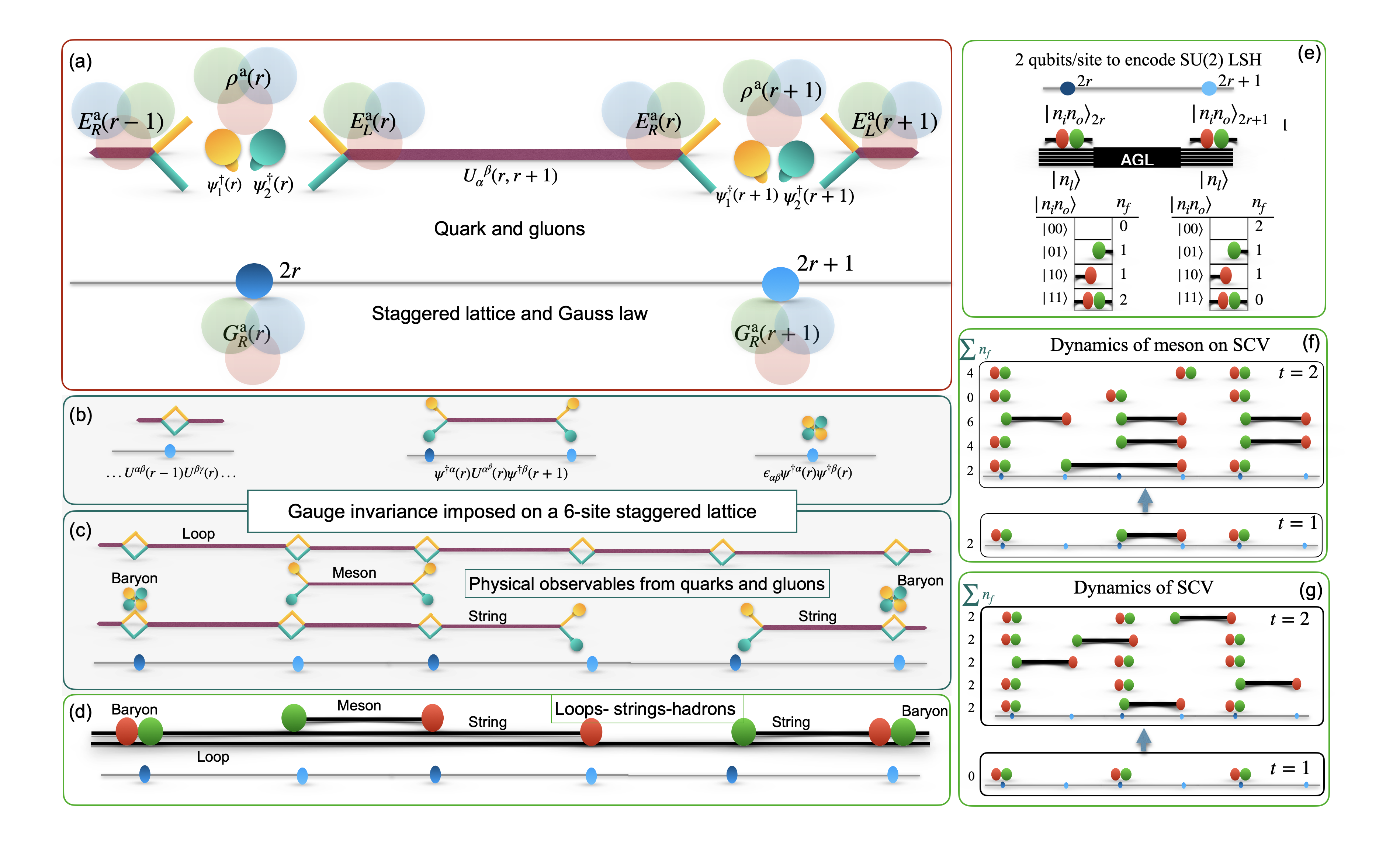}
    \caption{\small\textbf{From gauge redundancy to physicality: encoding into qubits} (a) A staggered lattice hosts quarks at even sites and anti-quarks at odd sites. The local electric fields ($E^{\mathrm a}$) and charge densities ($\rho^{\mathrm a}$) of a non-Abelian gauge theory carry color index $\mathrm a=1,2,3$, add up to form the Gauss law operator $G^{\mathrm a}$ that satisfy SU(2) Lie algebra with at each site. The link operators and matter fields carry fundamental group index $\alpha=1,2$ at each site, which transform by the generators of the Lie algebra present locally as $(*)^\alpha \rightarrow (G^{\mathrm a})^{\alpha}{}_\beta (*)^\beta$. we represent the variable of SU(2) gauge theory with  fundamental indices to contain open arms, and he ones carrying adjoint indices with tri-colored circles. (b) The physical states are defined to be gauge invariant, which are annihilated by $G^{\mathrm a}(r)$, for all $\mathrm a, r$. A cartoon representation of gauge-invariant objects is shown, with no open arms to any of the variables denoting all indices contracted. (c) In conventional gauge theory formulations, this construction involves forming singlets, which are non-local for loops and strings/mesons but on-site for Baryons. (d) The LSH framework directly maps these compound gauge singlets to loop-string-hadron (LSH) degrees of freedom, as shown in this panel. A global physical state is represented by a direct product state in LSH basis, characterized by only integer quantum numbers $\Pi_r|n_l,n_i,n_o\rangle_r$. Here $n
    _l$ counts number of flux lines passing, $n_i$ counts number of red balls and $n_o$ counts the number of green balls at each site. (e)  The onsite LSH basis is directly used as a computational basis. The true dynamical degrees of freedom are the string ends $n_i,n_o$ at each site. The local fermion number is defined on the staggered lattice as $n_f(r)= n_i(r)+n_o(r)$ for even sites and $n_f(r)= 2-[n_i(r)+n_o(r)]$ for odd sites. At any site, $n_f=2$ denotes the presence of a Baryon, $n_f=1$ denotes the presence of the end of a meson or a longer string, and $n_f=0$ denotes the vacuum for fermions. The strong coupling vacuum (SCV) is the state where all odd sites are fully filled and all even sites are empty, denoting $n_f=0$ for all sites. 
    (f,g) The diagrams illustrate dynamics starting from a single meson placed on top of SCV and the SCV itself. The LSH Hamiltonian dynamics buildup entanglement and cause particle number fluctuations, as illustrated just for a single Trotter steps under the LSH Hamiltonian. The system is initialized at $t=1$ in a zero-entanglement product states. At $t=2$, the evolution transitions the system to complex superpositions of states with varying total particle numbers ($n_f$), governed by the interplay of the electric ($H_E$) and mass ($H_M$) Hamiltonian terms. This fluctuation of total particle number is a hallmark of relativistic quantum field theory. The branching of the wavefunction depicted here highlights the rapid growth of entanglement, rendering the full 120-qubit simulation classically intractable. starting with the central meson state, the values of $n_f(r,t)$ across the lattice at all time steps are measured, and the same for time evoled SCV is also measured. Their difference is calculated to identify the propagation of the meson and is reported in the experimental heatmaps in Fig. \ref{fig:HGplot}.}
    \label{fig:dof-lsh}
\end{figure*}
In this work, we focus on the simplest continuous, yet non-Abelian, gauge group SU(2) in $(1+1)$ spacetime dimensions with the ultimate aim of simulating the strong interactions of nature, described by SU(3) gauge theory in $(3+1)$ dimensions. The system consists of dynamical fermions (matter) interacting via SU(2) gauge fields defined on the links of a spatial lattice. Understanding the structure and dynamics of entanglement entropy provides a new tool in the era of quantum information science \cite{Amorosso:2024leg}. Although various dynamical phenomena have been envisaged in different model-building and phenomenological predictions \cite{Altmann:2024kwx, Muller:2022htn}, the scientific community has been waiting for an ab initio calculation or a quantum simulation demonstrating its validity and revealing many quantum mechanisms underlying the existing effective or microscopic description of physics in out-of-equilibrium or in extreme environments.
\subsubsection*{The Hamiltonian and Its Continuum Limit}
The dynamics of gauge fields coupled with staggered fermions are governed by the Kogut-Susskind Hamiltonian \cite{Kogut:1974ag}. The Hamiltonian contains an electric energy term $(H_E)$, a staggered mass term $H_M$ and a matter gauge interaction term $(H_I)$ combined as:
\begin{eqnarray}
\label{eq:Hamiltonian}
    H =  \frac{g^2}{2a} H_E + m H_M + \frac{1}{2a} H_I 
\end{eqnarray}
The parameters $g$, $m$, and $a$ denote the gauge coupling, lattice fermion mass, and lattice spacing, respectively. The central challenge in simulating high-energy physics is to recover the \textit{continuum limit}, in which the lattice spacing $a \to 0$ and the discrete theory faithfully reproduces the continuous quantum field theory (QFT).  In $(1+1)$ dimensions, this limit corresponds to the regime of vanishing coupling, $g \to 0$, provided the simulation volume is sufficiently large to capture the relevant physics. The volume of the system is given by $L=Na$, for an $ N$-site system. The Hamiltonian given in (\ref{eq:Hamiltonian}) can be scaled as 
\begin{eqnarray}
    W=\frac{2a}{g^2}H = H_E+\mu H_M+xH_I 
    \label{eq:ham}
\end{eqnarray} 
The couplings in $W$ with electric, mass and matter-gauge interaction terms are dimensionless and given as $1, ~ \mu,~ x$ respectively, where $x= \frac{1}{g^2a^2} ~~\& ~~\mu=2\frac{m}{g}\sqrt{x}$. For a chosen fixed value of $\frac{m}{g}$, the continuum limit of the theory lies at $N\rightarrow \infty$ and $x\rightarrow \infty$ \cite{Hamer:1992ic}. The work primarily focuses on $N=60$ and $x=100$
, where the tensor-network and Pauli-propagation baselines remain sufficiently controlled to enable meaningful comparison with the QPU data within the chosen accuracy targets.
Within statistically comparable uncertainty, the QPU runtime remains set primarily by the fixed shot budget, whereas the available classical baselines show increasing computational cost.  
We further study a lower and a higher value of $x$, to examine how the quality and cost of the available classical baselines evolve as the weak-coupling regime is approached at fixed system size.  This is presented later in this article.

\subsubsection*{Encoding Non-Abelian Gauge Invariance}
A crucial hurdle in simulating the dynamics of a gauge theory arises from the need for gauge invariance. Although a gauge-invariant and orthonormal basis provides a natural choice for efficient encoding, commonly used basis - such as Wilson loops and strings - are inherently non-local. A further difficulty lies in expressing the Hamiltonian in terms of universal gate sets implementable on physical quantum hardware. When dealing with noise in today's quantum hardware, locality can be crucial for maintaining gauge invariance throughout the dynamic simulation. To date, a scalable and immediately implementable algorithm for non-Abelian gauge theories has remained absent, even in the era of utility-scale quantum hardware \cite{Kim:2023bwr}.  

 The dynamics, governed by the Hamiltonian in (\ref{eq:Hamiltonian}}), preserves gauge invariance generated by a set of three Gauss law operators $G^{\mathrm a} (r)$ for $\mathrm a=1,2,3$ and $\forall r$. These generators satisfy the SU(2) algebra at each lattice site. The physical or gauge invariant states are defined to be annihilated by all the SU(2) Gauss's law constraints at all lattice sites. In literature, these gauge-invariant degrees of freedom are described by the non-local Wilson loops, strings, and hadrons (mesons and baryons). 
 This work adopts the loop-string-hadron (LSH) framework for SU(2) lattice gauge theory \cite{Raychowdhury:2019iki} as an efficient approach that addresses  the challenges associated with a non-abelian gauge theory and non-local interactions.

 The LSH framework  is a reformulation of Kogut-Susskind's original framework \cite{Kogut:1974ag}, obtained via the prepotential formalism \cite{Mathur:2004kr, Mathur:2007nu, Anishetty:2014tta, Raychowdhury:2018tfj}.
 The LSH basis states can be intuitively understood to be on-site snapshots of all possible non-local loops of electric fluxes, strings connecting a quark matter, and hadrons (baryons and mesons), denoted as $|n_l,n_i,n_o\rangle_r$, that can be present globally on the lattice as denoted in Fig. \ref{fig:dof-lsh} and satisfy an Abelian Gauss Law (AGL) constraint on each link, ensuring the continuity of electric flux lines. The LSH Hamiltonian consists of the diagonal number operators and ladder operators for the local LSH degrees of freedom and satisfies all the AGLs.

The SU(2) LSH Hamiltonian in $1+1$ dimension has been extensively analyzed in the context of developing efficient quantum algorithms for simulating this theory and it has shown to offer significant reduction in qubit costs and gate depth, for both near and far-term quantum hardware \cite{Davoudi:2022xmb}. The Hamiltonian is expressented in terms of occupation number and ladder operator for the on-site LSH degrees of freedom. 

In contrast, this work presents quantum simulation of LSH dynamics implemented on state-of-the-art 156-qubit IBM quantum processor. The aim of the work is to probe the continuum limit, which requires probing the dynamics in the weak-coupling regime $x>>1$ for a lattice of reasonably large size.  
In this parameter regime, (i) the dynamics is dominated by the off-diagonal term in (\ref{eq:ham}). For an approximation performed for $H_E$, it affects the dynamics less significantly, as compared to the other coupling regime $x<1$.
(ii) Without any loss of generality, one may consider a lattice with open boundary conditions, where a large number of incoming fluxes just pass through the lattice keeping the global charge sector same as the strong coupling vacuum (SCV)\footnote{\label{fn-1} The symmetry of LSH Hamiltonian preserves the following global charges  \\
 (i)  $ \mbox{total Baryon number: } \mathcal B = \sum_{r=0}^{N-1} (n_i(r)+n_o(r)) -N $, and (ii) $ \mbox{ net lattice flux: } q = \sum_{r=0}^{N-1} (n_o(r)-n_i(r))  $. This translate to conserving the global observables $q_o=\sum_{r=0}^{N-1} n_o(r)$ and $ q_i=\sum_{r=0}^{N-1} n_i(r)$.
}. The contribution of this background fluxes can be approximated as a global phase in this regime. This weak coupling approximate version of the LSH Hamiltonian has previously been developed in the context of analog simulation \cite{Dasgupta:2020itb}. 
The algorithm identifies certain fermionic configurations at a site $(n_i=0,n_o=1)$, when a new flux is created on a link emerging from the lattice, and its electric energy contribution is counted to contribute.  


\subsubsection*{Encoding and preparing the initial states}
\begin{figure*}[t!]
    \centering
    \includegraphics[width=1
    \linewidth]{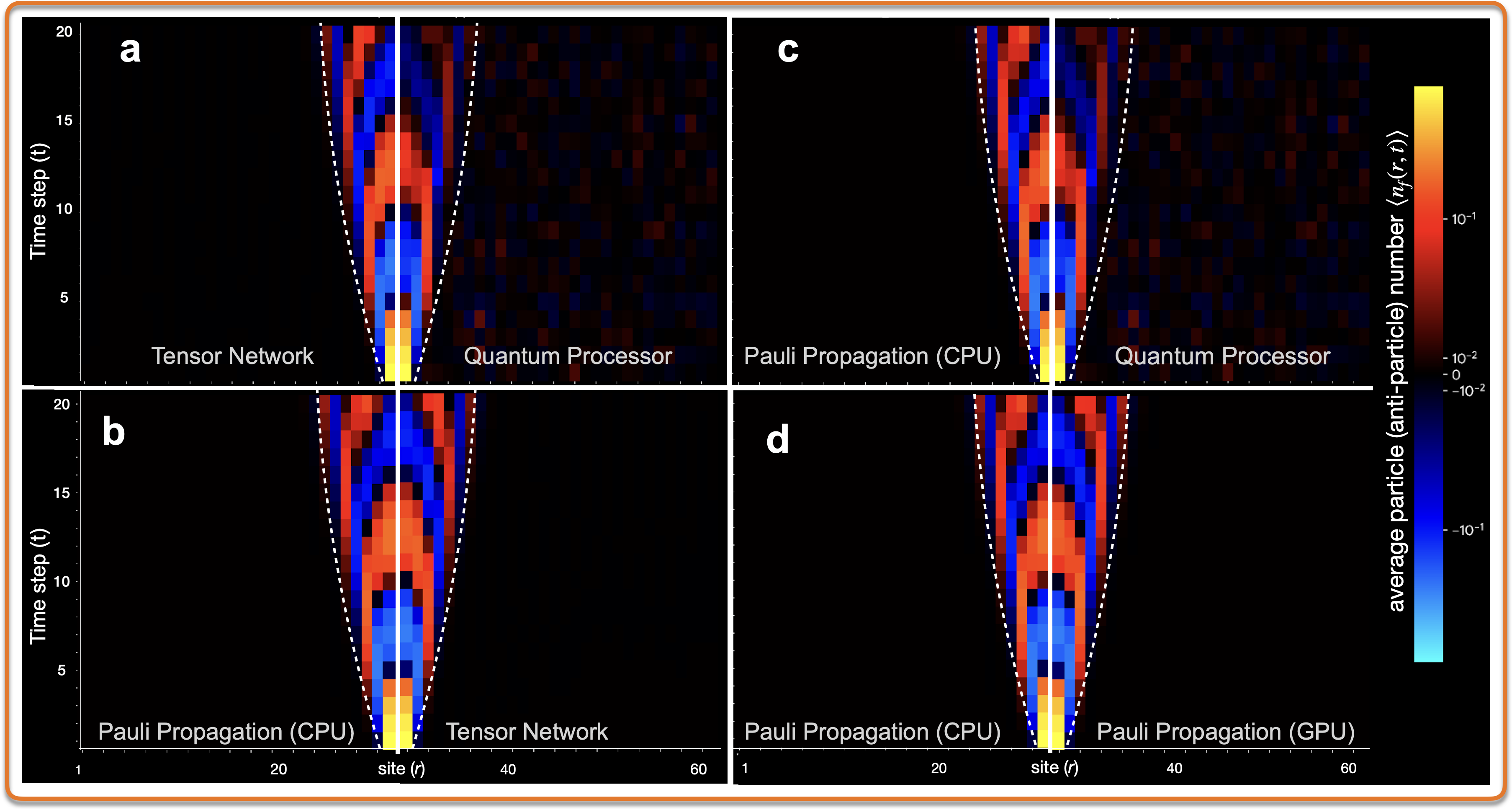}
    \caption{\small\textbf{Real Time Propagation of SU(2) Hadron: high fidelity validation of the ansatz, algorithm, and hardware implementation.}  Space-time evolution of quark (anti-quark) number $n_f(r,t)$, starting from an initial meson placed at the centre of the lattice, is obtained on a 60-site lattice.  The heatmaps show internal dynamics within the hadron caused by dynamical phenomena such as pair creation and annihilation, mesonic string breaking and rejoining. A dotted line is traced on the heatmaps (a)-(d) to guide the eye to identify the meson's confined structure. Internal oscillations are observed as periodic red and blue regions within the meson boundary. The same result is obtained using four different approaches: (i) Quantum simulation on superconducting quantum processor (QPU) \textit{IBM Boston}; (ii) Tensor network (TN) calculation for the full LSH Hamiltonian; (iii) Pauli Propagation (PP) on GPU that considers a minimum threshold of $10^{-5}$ for the coefficients of the Pauli terms, and (iv) PP with a truncation in the maximum number of allowed terms in the operator expansion on a CPU. 
    The dynamics on the left half (lattice sites $0-29$) of the figures (a)-(d)  are from one calculation, while the right half of the plot shows another.
  (a) The match between the left and right halves of the heatmap confirms the validity of the experimental result as a simulation of the full SU(2) gauge theory. 
  The TN calculation is performed for the full LSH Hamiltonian, with time evolution obtained by 2-site TDVP with a maximum bond dimension of $200$, and doesn't involve Trotterization error. The Quantum simulation is performed using the weak-coupling approximated LSH and first-order Trotterization. The match thus validates the experimental observation to be a simulation of the actual theory. 
  (b)  PP provides a classical simulation of the noiseless quantum circuit. An exact match with the TN calculation done for the full theory validates the quantum algorithm's ability to capture the dynamics of the original SU(2) theory.
  (c) The mismatch between PP and QPU identifies the effect of hardware noise in experimental observation. (d) PP, performed on the CPU, and the GPU match identically, at least for early time, and validate the two different truncation schemes. }
    \label{fig:HGplot}
\end{figure*}

 In a $1+1$-dimensional lattice, electric flux loops can flow in only one direction, and at each site a fermionic doublet can exist. This translates to an on-site physical state at $r$ to be characterized as 
$|n_l (r),n_i(r),n_o(r)\rangle.$
As illustrated in Fig. \ref{fig:dof-lsh}, the manifestly gauge singlet quantum numbers, $n_l$, denotes the electric flux passing through the site without any change and can vary from zero to infinity; $n_i=1$, denotes an incoming flux being absorbed at that site by an on-site fermion forming a string-end like object for an incoming string; $n_o=1$, denotes an outgoing flux being created at that site by an on-site fermion forming a string-end like object for an outgoing string. Fermionic statistics restricts $n_i,n_o$ to take values between $(0,1)$, while being bosonic, $n_l$ can be any positive semi-definite integer. Abelian weaving across the neighboring sites following the AGL is given by the on-link constraint
 \begin{eqnarray}
    \mathrm{Abelian~Gauss~Law:}\,
    n_l+n_o(1-n_i)\Big|_r= n_l+n_i(1-n_o)\Big|_{r+1}
 \end{eqnarray}
 As a consequence of the fact that gauge field are not dynamical for one spatial dimension, the quantum number $n_l$ across the lattice is determined by the boundary flux and the fermion configurations throughout the lattice in order to satisfy the AGL at all the links of the lattice.
This basis, being a strong coupling basis, yields the electric and mass part of the Hamiltonian $H_E, H_M$ to be diagonal terms.  While the off-diagonal terms of the Hamiltonian, the matter-gauge interaction $H_I$, cause the dynamics of loops, strings and hadrons on the lattice. 

\begin{figure*}[t!]
    \centering
\includegraphics[width=1\linewidth]{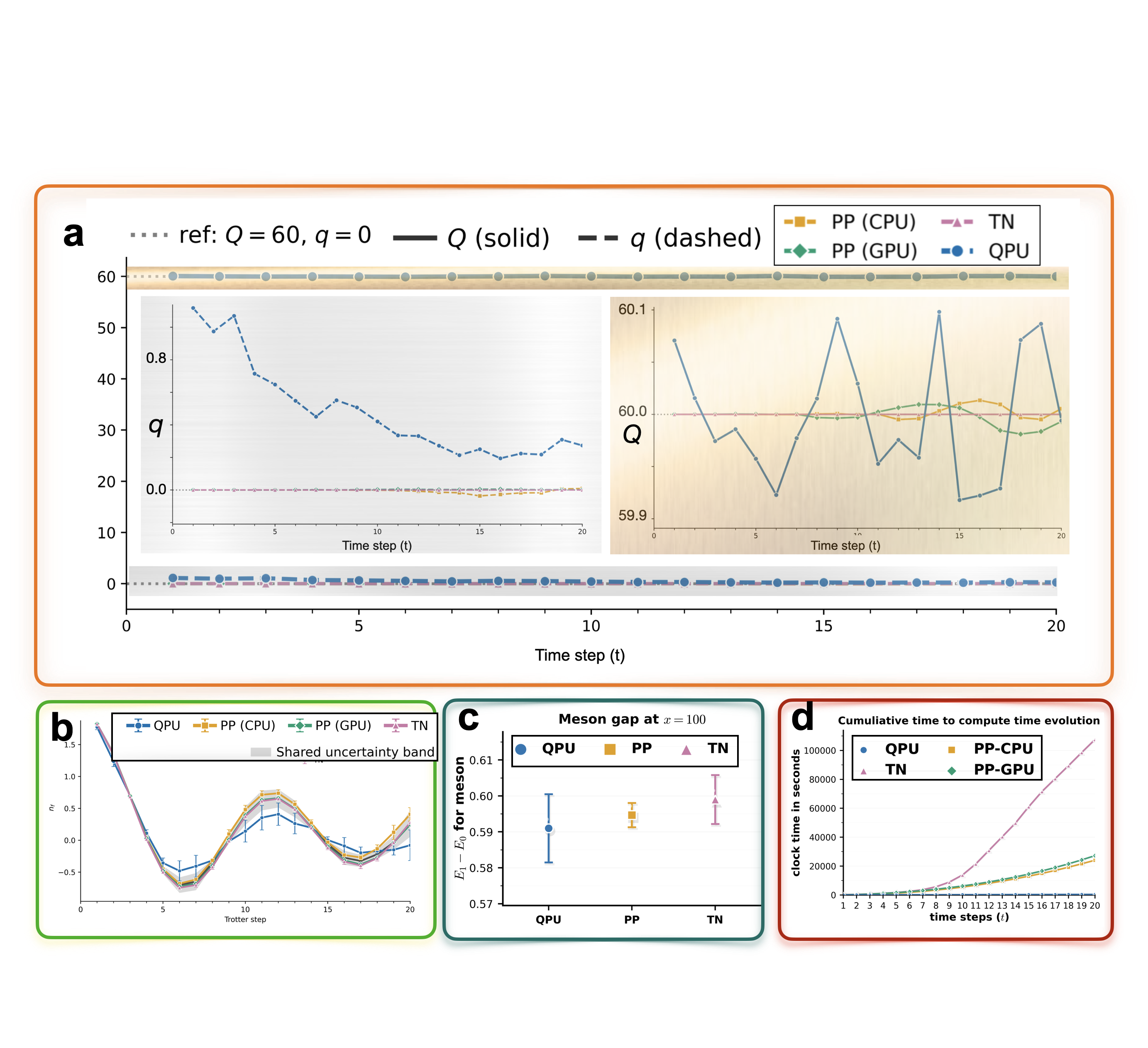}
    \caption{\small{\textbf{ Results: Symmetry protection, error bound, compute time, and empirical hadron spectrosopy}} (a) Global-symmetry conservations through the total charge $\mathcal B+N=Q(=60)=\sum_r (n_i(r)+n_o(r))$ and $q= \sum_r (n_i(r)-n_o(r))$ is monitored for all four methods. This directly implies  preserving all the gauge symmetries in LSH dynamics \cite{mathew2022protecting}. Despite hardware noise, the QPU maintains symmetry within $0.33\%$ for $Q$ and $2\%$ for $q$. 
    (b) Observable corresponding to average particle-antiparticle density $n_f(t)=\sum_r n_f(r,t)$ is plotted as function of time and depicts regular oscillation with decreasing amplitude, denoting the notion of thermalization. This global observable is obtained for each method,
 the median is used as the reference point, and the intrinsic method-to-method uncertainty is taken as the sample standard deviation.
Per site contribution of measuring global charge $Q$ contributes to $n_f$, hence its per-site uncertainty is added to get a final shared uncertainty band. Additionally, for each method, we report individual error bars reflecting deviations from the median. The error bars thus capture both cross-method variability and global-constraint inconsistency.   The error bars are statistically comparable and are correlated with the computation time for each method. (c) The frequency of breathing mode oscillation $\omega$ of the meson is calculated using oscillation only at the central part of the lattice, across the mesonic length, and also for the variable $n_F$ and are found to be consistent. Here frequency is reported using the first 15 time step data (for all methods) for $n_F$ and fitting the same in a stretched damped oscillation function. This frequency corresponds to the energy gap of the first excitation energy $E_1$ of the meson from its initial state $E_0$.(d) The cumulative clock time taken for each method to compute dynamics upto certain Trotter step. QPU time for a fixed budget of $10k$ shots per trotter step was a constant 20 seconds per trotter step; the time complexity for both the classical methods, PP and TN, increased exponentially, the latter being much faster.  }
    \label{fig:analysis100}
\end{figure*}
\section*{Observing Real-time Dynamics} 

The initial state is chosen to be an unentangled state, which is a strong coupling eigenstate in LSH basis. A natural choice of such a state is the SCV, which corresponds to a `no particle - no antiparticle' state with $n_f=0$ at all sites. Next, a different initial state is prepared, where a meson is placed at the middle of the lattice, where $(n_f(r)=1, r=N/2-1, N/2)$, while keeping $n_f=0$, elsewhere as illustrated for a small lattice in Fig. \ref{fig:dof-lsh}.

\subsection*{Validation hierarchy: theory, approximation, circuit, hardware}
We study the quench dynamics of a single meson propagating through a dynamical medium on a 60-site staggered lattice for up to 25 Trotter steps. For each evolution time $t_n$, we prepare and evolve two initial states on the QPU: the strong-coupling vacuum (SCV), and a state in which a meson is placed at the centre of the lattice on top of the SCV. In both cases, the qubits are measured in the 
$\sigma_z$ basis, and the corresponding probabilities are estimated from repeated experimental runs. This procedure is repeated for $t_n=1,2,3,\ldots,25$ time steps. A schematic representation of the protocol is shown in Fig. \ref{fig:dof-lsh} (f) and (g). 

The observable used throughout this work is obtained by subtracting the probability distribution measured for the evolved SCV from that measured for the evolved mesonic state. This differential protocol suppresses background contributions and isolates the coherent signal associated with meson dynamics. The resulting space-time evolution is shown in Fig. \ref{fig:HGplot}.
The QPU data, together with the classical baselines, reveal an early-time oscillatory mode within the mesonic profile in addition to the overall confined propagation. This oscillatory structure motivates the spectroscopic analysis presented below. In the present work, however, our emphasis remains on the experimentally accessible early-time regime, where the layered comparison between theory, approximation, circuit and hardware can be carried out in a controlled way.

The experiment on the QPU is performed with a circuit containing more than $ 17000$ two-qubit gates and $ 90000$ single-qubit gates, and a fixed $ 10,000$-shot budget per Trotter step. The depth of 2-qubit gates for simulating the Time evolution grows up to $259$ for $20$,  and to $324$ for $25$ Trotter steps, respectively. Using only a low-cost readout error mitigation \cite{Berg:2020ibi}, the signal from QPU shows a coherent signal for propagation of a hadron by extending its size at each Trotter step, but confined within a `light-cone'. The edges of the `light cone' trace a curved path instead of a straight line, denoting the constituent quarks to stay confined by the strong force instead of flying away freely. 

The experimental observation is benchmarked by two classical computing methods: (i) Tensor Network (TN)  calculations performed for the original LSH Hamiltonian and basis, via construction of Matrix Product Operators (MPO) and Matrix Product States (MPS) respectively, carefully created as a toolbox for LSH calculations \cite{Mathew:2025fim, Gupta:2026tcg}. The time evolution is computed using a 2-site TDVP algorithm that is free of Trotterization error. The bosonic cut-off for this computation is set at $2j_{max}=5$, while a cut-off in bond dimension is set at $D_{max}=200$. 
(ii) The Pauli Propagation (PP) method  \cite{Rudolph:2025gyq} is used to simulate the noiseless digital circuit, with truncation introduced either through a cap on the number of retained Pauli terms or through a coefficient threshold. These two baselines serve different roles: TN provides a reference for the physical dynamics of the full theory over the accessible time window, whereas PP validates the digital implementation of the approximated dynamics.

Fig. \ref{fig:HGplot} 
summarizes this validation hierarchy. Agreement between the TN and PP results supports the weak-coupling approximation and the digital algorithm over the accessible time window. Comparison between PP and QPU then isolates the effect of hardware noise on the experimentally observed signal. Across these comparisons, the main qualitative features, the confined propagation pattern and the internal oscillatory structure of the mesonic profile, remain robust.

\subsection*{Symmetry protection and robustness}

Because exact dynamical results are not available for the full system size studied here, we quantify the quality of the observed particle-density dynamics $n_f(t)=\sum_{r=1}^{60}n_f(r,t)$ using two complementary diagnostics: (i) the spread across methods and (ii) the deviation from exactly known conserved global quantities. Globally conserved charges as per \ref{fn-1} are exactly known. For the chosen initial states, these should take the value $Q=60,q=0$ throughout the dynamics. The observable $n_f(r,t)$ is directly related to the on-site contribution to the global charge $Q$. These diagnostics provide a practical error estimate for the experimentally observed signal within the accessible time window.


The global charges $Q= \sum_{r=1}^{60 } n_o(r)+n_i(r) ~\& ~ q= \sum_{r=1}^{60 } n_o(r)-n_i(r)$ remain close to their expected values throughout the observed dynamics as shown in Fig. \ref{fig:analysis100}, providing a sensitive indication that the evolution remains within the intended physical sector \cite{mathew2022protecting}. The corresponding $n_f(t)$ from QPU, TN and PP are statistically comparable within the reported uncertainty bands, while the computational cost of the classical baselines increases substantially over the same time window. For the tensor-network calculations, this deterioration is tied to the finite bond-dimension budget used in practice ; for PP, it reflects the growing cost of maintaining an accurate truncated Pauli expansion.

 The tensor network experiment on classical computing hardware encounters an entanglement wall. The time complexity grows exponentially with the entanglement entropy. Ideally, it requires increasing the maximum allowed bond dimension $D_{max}$ to minimise the error. In this work, $D_{max}$ was limited to $200$ due to constraints on available computing resources. With this particular limit on $D_{max}$, the dynamics beyond 20 Trotter steps are obtained to contain errors which can not be neglected and hence are not reported. The details are reported as supplementary information. 
 In summary, the available classical baselines become increasingly difficult to control as the accessible time window is extended at fixed computational budget, whereas the runtime on QPU remains set primarily by the shot budget. Within this experimentally accessible regime, the differential measurement protocol yields observables that remain robust against hardware noise. Invading long-time dynamics for $x\rightarrow \infty$ will ultimately require improved control over hardware noise and coherence. The reported dynamics of the observable exhibit noise resilience under the differential measurement protocol employed in this experiment. 
 \begin{figure*}[t!]
    \centering
    \includegraphics[width=1\linewidth]{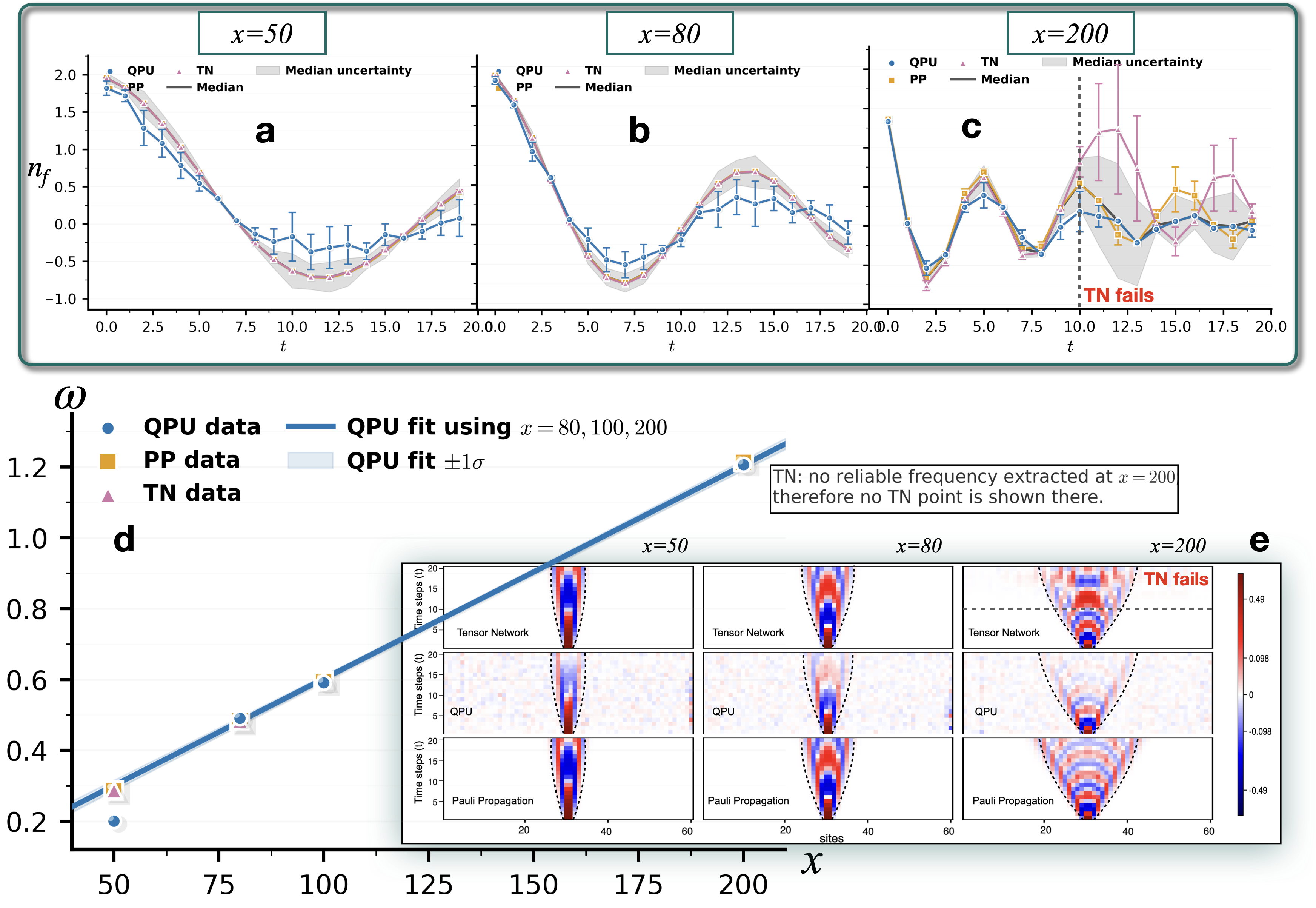}
    \caption{\small{\textbf{Demonstrating robust quantum simulation for varying coupling $x$. }} 
    (a)-(c) Oscillation in total particle number obtained by individual methods are displayed along with error bar calculated as done in Fig. \ref{fig:analysis100}. 
    (d) The frequency of breathing mode oscillation is calculated for $x=50, 80, 200$ as done for $x=100$ in Fig. \ref{fig:analysis100} and plotted (together with the value for $x=100$) against $x$. For each of the individual methods, the fit is linear. Note that,the frequency obtained for $x=50$, for QPU is significantly lower than other methods, and is not included in the fitting for QPU. (e) High-fidelity dynamics is demonstrated via heatmaps of data obtained from quantum measurements in the QPU and from classical simulations using TN (for full LSH) and PP (for the quantum circuit).  From (c) and (e), it is clearly found that TN breaks down for $x=200$, after the $10^{\mbox{th}}$ time step. Consequently, no frequency is obtained by fitting this data in a stretched damped oscillation fit displayed in (d).  }
    \label{fig:comp-x}
\end{figure*}
 
\subsection*{Breathing-mode spectroscopy and coupling dependence}
 A central physical observable in this work is the early-time internal oscillation of the mesonic profile. We extract a characteristic frequency from three related quantities: (i) oscillation in particle density at the central two sites of the lattice (ii) second moment of particle density distribution inside the confined meson and (iii) frequency of the fluctuation in $n_f$ as reported in Fig. \ref{fig:analysis100}. The frequency value obtained via these three methods remains consistent while the fluctuation in total number density is found to be dominated by the fluctuation at the central sites. We therefore interpret the resulting frequency as a spectroscopic proxy for the lowest excitation gap of the mesonic configuration above its initial state, $(E_1-E_0)$. The frequencies are obtained with an error bar by fitting the data into a stretched damped oscillatory function as discussed in Methods. 

The robustness of the quantum simulation protocol is further tested for different values of the coupling $x =50, 80, 200$, without changing the value of $m/g$, in order to study how the observed dynamics and the quality of the classical baselines evolve with coupling. The hardware data show the expected narrowing or widening of the confined propagation pattern across this range, while TN and PP provide corresponding reference calculations wherever they remain controlled. These results are summarized in Fig. \ref{fig:comp-x}. Clear signature of internal oscillations are present at all couplings, thereby demonstrating the robustness of the algorithm and its implementability on a QPU.
 
 The coupling dependence of this frequency provides an additional consistency check. If the initial mesonic object is characterized by a length scale $L\sim ga=\frac{1}{\sqrt{x}}$, then a simple confined-particle picture suggests a gap scale proportional to $\frac{1}{L^2}$ or proportional to $x$. Interestingly, Fig. \ref{fig:comp-x} demonstrates a linear relation between the breathing mode frequency $\omega$ and $x$ as obtained by QPU (and also using PP and TN data). The breathing mode frequency obtained from QPU for $x=80,100,200$ fits perfectly to a straight line with slope $0.00602299\pm 0.00016134$ when plotted against x, and the PP and TN data (whenever accessible) perfectly lie on this straight line as shown in Fig. \ref{fig:comp-x}. The error obtained in fitting the QPU data is only $2\%$.

 \subsection*{ Limits of current classical baselines}

For the QPU measurements, the runtime per time step is set primarily by the fixed shot budget. For PP, the runtime depends on the truncation strategy used to control the Pauli expansion, while for TN it depends on the bond-dimension resources required to maintain accuracy. These differences are relevant when comparing the practical reach of the available baselines. The Pauli propagation method \cite{Rudolph:2025gyq} provides an efficient classical algorithm for simulating an error-free quantum circuit. This method is employed on the CPU, with an estimate of the max-term obtained by extrapolating exact results from small systems to minimise truncation error. We further employ another strategy, PP \emph{on GPU}, in which terms with coefficients exceeding a threshold are considered. In summary, the PP calculations provide a useful circuit-level baseline for the noiseless digital dynamics, but their accuracy and cost depend on the truncation scheme used.

Although the circuit depth is held fixed across these coupling values, the cost and reliability of the classical baselines change significantly. In particular, the TN calculations become increasingly difficult to converge as $x$ is increased, and for 
$x=200$, they cease to provide a reliable reference beyond the early-time window shown. The PP calculations also show increasing sensitivity to truncation at larger 
$x$, as discussed in the Supplementary Information. These observations indicate that the available classical approximations become increasingly strained as the weak-coupling regime is approached at fixed system size.

In summary, the coupling scan shows that the experimentally observed mesonic signal remains robust across the values of $x$ studied here, while the available classical baselines become progressively harder to control within fixed resource budgets, and provides an empirical indication that the tested TN and PP strategies face increasing practical limitations in the same regime where the hardware measurements continue to yield a coherent differential signal. The extracted oscillation frequency remains sufficiently robust across the explored couplings to support a coupling-dependent spectroscopic analysis. For 
$x=50, 80,100,200$, the QPU values follow an approximately linear trend, with PP and TN lying close to the same behaviour wherever reliable baseline data are available. At 
$x=200$, the deterioration of the TN reference is itself evident in Fig. \ref{fig:comp-x}, reinforcing that the coupling scan should be interpreted jointly as a spectroscopy result and as a practical comparison of currently available baselines.
 
\section*{Outlook}

The current work paves the way for performing large scale ab initio calculations for non-Abelian gauge theories using a quantum computer, and demonstrates that scalable and useful physical results, such as the breathing mode frequency of hadrons (or the energy gaps) are reliably computed even while using a noisy quantum hardware. By successfully accessing a 60-site lattice, the current work establishes that pre-fault-tolerant hardware, when combined with hardware-efficient encodings such as LSH and error-robust measurement protocols, can capture non-perturbative dynamics inaccessible to classical methods. 

Three immediate directions follow from this result.
First, extending the simulation time beyond the current window is an immediate priority. Our results hint at the early stages of hadron breathing, pair production and string breaking dynamics within a confined meson; prolonging the evolution would allow direct observation of complete hadronization of the vacuum, a phenomenon that is central to understanding the thermalization of the quark-gluon plasma formed in heavy-ion collisions 
\cite{Muller:2011ra,Muller:2022htn,Altmann:2024kwx}.

Second, the LSH framework utilized here is naturally extensible to higher dimensions and more complex gauge groups. The most significant barrier to simulating Quantum Chromodynamics (QCD) has been the implementation of the SU(3) gauge group. The methods validated in this work for SU(2) provide a blueprint for advantage in encoding SU(3) gauge theory to qubits via the LSH framework built upon the prepotential formulation \cite{Raychowdhury:2013rwa}. The 1+1-dimensional framework is ready to use \cite{PhysRevD.107.094513}, while the higher-dimensional framework is being developed step by step \cite{PhysRevD.111.074516, kadam2025loopstringhadronapproachsu3lattice}.

Finally, the efficient encoding and the error-resilient differential observation technique used in this work, along with their confirmation of validity in capturing correct physical dynamics, establish a pathway for computing novel real-time dynamics with quantum advantage \cite{lanes2025frameworkquantumadvantage}. 
We demonstrate that our differential measurement protocol effectively isolates the coherent physical signal even without an active error mitigation. As hardware fidelity continues to improve, along with appropriate active error mitigation or early error-correction protocols, these methods are expected to enable increasingly sophisticated simulations of scattering phenomena. When applied to finite-density systems where the Monte Carlo sign problem is most severe, it could soon unravel the phase diagram of dense nuclear matter, shedding light on the interiors of neutron stars and the early universe. 

Furthermore, quantum simulation experiments in future 
can lead to the ability to study collisions of mesons and baryons and to probe parton distribution functions and deep-inelastic scattering processes from first principles, which would mark a substantial expansion in the range of non-equilibrium quantum field theory calculations accessible to computation, extending beyond what is currently practical with classical lattice QCD techniques.

\section*{Data Availability}
The data supporting the findings of this study are available at the public \href{https://github.com/mathew0036/lsh_data}{GitHub}  repository. 
\section*{Code Availability}
The codes used for the findings of this study are available from the corresponding author upon reasonable request.

\section*{Acknowledgments}
The authors would like to thank Abhinav Kandala and Nick Bronn for insightful discussions and feedback at various points of this work and for sharing comments on the manuscript. We would also like to thank Rudranil Basu for discussion and help in GPU implementation of the Pauli propagation method.
IR would like to thank Jesse Stryker, Zohreh Davoudi, Saurabh Kadam, and Navya Gupta for their continuous contributions in developing the loop-string-hadron framework and for numerous discussions during all collaboration meetings.
The work reported here is performed using the IBM Quantum Credits received by IR. The authors acknowledge the support. IR acknowledges useful discussions during meetings of QC4HEP working group.
Research of IR is supported by the  OPERA award (FR/SCM/11-Dec-2020/PHY) from BITS-Pilani and the cross-discipline research fund (C1/23/185) from BITS-Pilani. FI is supported by the cross-discipline research fund (C1/23/185) from BITS-Pilani. MdOA acknowledges computational and other support from BITS Pilani through another cross-discipline research fund (C2/24/282) from BITS-Pilani. 

\section*{Authors' contributions}

\textbf{FI} developed the quantum algorithm and benchmarked it against exact diagonalization, prepared the POC, performed noiseless classical simulation of the quantum algorithm.\\
\textbf{RM} helped design and map the quantum circuit tailored to the hardware, and ran all the quantum experiments, and performed noiseless classical simulation of the quantum algorithm.\\
\textbf{EM} performed the tensor network calculations to benchmark the performance of QPU, making all the plots and analyzing data.\\
\textbf{MdOA} set up the code in CUDA  and performed the classical simulation using GPU.\\
\textbf{NE} helped analyze the quantum circuits and experimental results \& evaluation, and contributed to manuscript preparation.\\
\textbf{IR} developed the problem statement, set up the algorithmic strategy, analyzed data, prepared the main manuscript, including graphics.\\

\section*{Competing interests}
The authors declare no competing interests.

\bibliography{references.bib}
\bibliographystyle{naturemag}


\newpage

{\Huge{\textbf{Methods}}}

\section*{The theory to be simulated}

In this section, we briefly review the theory we aim to simulate, the Kogut-Susskind Hamiltonian formalism, represented in the loop-string-hadron basis. As discussed in the main text, the continuum limit weak-coupling regime of the theory, encoded in the gates of the current experiment.
\subsection*{Kogut-Susskind Hamiltonian}
\label{sec: KSH}

The scaled Kogut-Susskind (KS) Hamiltonian describing SU(2) Yang Mills theory coupled to staggered fermions on $(1+1)$-d (1d spatial lattice and continuous time) \cite{Kogut:1974ag} can be written as:
\begin{eqnarray}
W&=&  H^{({\rm KS})}_E+ \mu H_M+ x H^{({\rm KS})}_I.
\label{eq:HKS}
\end{eqnarray}
 The gauge link $U(r)$ is a $2\times2$ unitary matrix defined on the link connecting  sites $r$  and $r+1$. A temporal gauge is chosen to derive the above Hamiltonian, which sets the gauge link along the temporal direction equal to unity.
The color electric fields $E_{L/R}^{a}$ are defined at the left $L$ and right $R$ sides of each link, and they satisfy the following commutation relations (SU(2) algebra) at each end:
\begin{eqnarray}
[E_L^a(r),E_L^b(r')]&=&i\epsilon^{abc}\delta_{rr'} E_L^c(r),
\nonumber\\
{[E_R^a(r),E_R^b(r')]}&=&i\epsilon^{abc} \delta_{rr'} E_R^c(r'),
\nonumber\\
{[E_L^a(r),E_R^b(r')]}&=&0,
\label{eq:ERELcomm}
\end{eqnarray}
where $\epsilon^{abc}$ is the Levi-Civita symbol.  The electric fields and the gauge link satisfy the following quantization conditions at each site,
\begin{eqnarray}
[E_L^a(r),U(r')]=-\frac{\sigma^a}{2}\delta_{rr'}U(r),
\nonumber\\
{[E_R^a(r),U(r')]}= U(r)\delta_{rr'}\frac{\sigma^a}{2},
\label{eq:EUcomm}
\end{eqnarray}
where $\sigma^a$ are the Pauli matrices. For a theory including matter fields, staggered fermionic fields $\psi^{\dagger\alpha} (r)$, for $\alpha=1,2$ are present at each lattice site.
The Hamiltonian in (\ref{eq:HKS}) is gauge invariant as it commutes with the  Gauss' law operator,
\begin{equation}
G^a(r)=E^a_L(r)+E^a_R(r-1)+\psi^\dagger(r) \frac{\sigma^a}{2} \psi(r)
\label{eq:Ga}
\end{equation}
at each site $r$.  The physical sector of the Hilbert space corresponds to the space consisting of states annihilated by (\ref{eq:Ga}). Solving the non-Abelian Gauss laws at each site $r$ as given in (\ref{eq:Ga}) is non-trivial, and engineering the same in an experiment is the most difficult job.

\subsection*{LSH Hamiltonian}
\label{sec:methods_LSHHamiltonian}
LSH formalism of lattice gauge theory is based on prepotential framework, where, the original canonical conjugate variables of the theory, i.e color electric field and link operators are replaced by a set of harmonic oscillator doublets, defined at each end of a link \cite{Mathur:2004kr,Mathur:2007nu,Mathur:2010wc,Anishetty:2009ai,Anishetty:2009nh,Anishetty:2014tta,Raychowdhury:2013rwa,Raychowdhury:2014eta,Raychowdhury:2018tfj}. In prepotential framework, the SU(2) gauge group is confined to each lattice site allowing one to have local gauge invariant operators and states at each site, leading to a description of local loop segments. Combining prepotentials with staggered fermionic matter fields at each lattice site to form gauge invariant singlets, yields on-site string-end operators. Staggered matter fields also combine into local gauge-invariant configurations representing hadrons, likewise in the original understanding of the theory. 
Thus the gauge invariant and orthonormal LSH basis is characterized by a set of three integers $
n_l(r), n_i(r), n_o(r)$, corresponding to the loop, incoming string, and outgoing string at each site (on-site hadron is equivalent to simultaneous presence of both the string ends). 
LSH basis states satisfy all the Gauss' law constraints by the construction. The allowed values of the LSH quantum numbers are
\bea
&&  0\le n_l(r)\le \infty , ~~ 0\le n_i(r)\le 1,~~0\le n_o(r)\le 1, ~~~
\eea
denoting $n_l$ to be a  bosonic excitation,  whereas $n_i,n_o$ to be fermionic in nature, even though the string ends contain information of both gauge field and matter content. 

A set of LSH operators consisting of both diagonal and ladder operators are defined locally at each site as:
\bea
 \hat n_{l/i/o}|n_l, n_i, n_o\rangle &=& n_{l/i/o}|n_l, n_i, n_o\rangle \\
\label{nlpm}
\hat \lambda^{\pm}|n_l, n_i, n_o\rangle &=& |n_l\pm 1, n_i, n_o\rangle \\
\label{nip}
\hat \chi_{i/o}^{+}|n_l, n_i, n_o\rangle &=& (1-\delta_{n_{i/o},1})|n_l, n_{i/o}+ 1, n_{o/i}\rangle \\
\hat \chi_{i/o}^{-}|n_l, n_i, n_o\rangle &=& (1-\delta_{n_{i/o},0})|n_l, n_{i/o}- 1, n_{o/i}\rangle 
\eea
The LSH basis states must satisfy the AGL, as explained in the main text to be counted as a valid loop-string-hadron segment of the global loop configurations present on the lattice. 

Hamiltonian of the theory, \bea
\label{eq:HLSH_full}
W^{(\rm LSH)}=H^{(\rm LSH)}_E+\mu H^{(\rm LSH)}_M+xH^{(\rm LSH)}_I
\eea
is exactly equivalent to the original Hamiltonian (\ref{eq:HKS}), with the following form of electric term, mass term and matter-gauge interaction term in terms of LSH operators:
\bea
\label{HELSH}
H^{(\rm LSH)}_E&=&\sum_{r=0}^{N-2}\Bigg[ \frac{\hat n_l(r)+\hat n_o(r)(1-\hat n_i(r))}{2},\nonumber \\
&& \times \left( \frac{\hat n_l(r)+\hat n_o(r)(1-\hat n_i(r))}{2}+1 \right) \Bigg]\\
\label{HMLSH}
H^{(\rm LSH)}_M &=& \sum_{r=0}^{N-1} (-1)^r(\hat n_i(r)+\hat n_o(r)), \\
\label{HILSH}
H^{(\rm LSH)}_I &=&\sum_{r=0}^{N-1} \frac{1}{\sqrt{\hat n_l(r)+\hat n_o(r)(1-\hat n_i(r))+1}}\times \\ \nonumber &&  \Big[ S_o^{++}(r)S_i^{+-}(r+1)
+ S_o^{--}(r)S_i^{-+}(r+1)\\ \nonumber &&
+S_o^{+-}(r)S_i^{--}(r+1)
+S_o^{-+}(r)S_i^{++}(r+1)\Big]\\ \nonumber & \times& 
\frac{1}{\sqrt{ \hat n_l(r+1)+\hat n_i(r+1)(1-\hat n_o(r+1))+1}}.
\eea
Here (\ref{HILSH}) contains LSH ladder operators in the following combinations (suppressing the explicit site index), 
\bea
S_o^{++}&=& \hat \chi_o^+ (\hat \lambda^+)^{\hat n_i}\sqrt{\hat n_l+2-\hat n_i} \\
S_o^{--}&=& \hat \chi_o^- (\hat \lambda^-)^{\hat n_i}\sqrt{\hat n_l+2(1-\hat n_i)} \\
S_o^{+-}&=& \hat \chi_i^+ (\hat \lambda^-)^{1-\hat n_o}\sqrt{\hat n_l+2\hat n_o} \\
S_o^{-+}&=& \hat \chi_i^- (\hat \lambda^+)^{1-\hat n_o}\sqrt{\hat n_l+1+\hat n_o)} 
\eea
and
\bea
S_i^{+-}&=& \hat \chi_o^- (\hat \lambda^+)^{1-\hat n_i}\sqrt{\hat n_l+1+\hat n_i)}  \\
S_i^{-+}&=& \hat \chi_o^+ (\hat \lambda^-)^{1-\hat n_i}\sqrt{\hat n_l+2\hat n_i} \\
S_i^{--}&=& \hat \chi_i^- (\hat \lambda^-)^{\hat n_o}\sqrt{\hat n_l+2(1-\hat n_o)} \\
S_i^{++}&=& \hat \chi_i^+ (\hat \lambda^+)^{\hat n_o}\sqrt{\hat n_l+2-\hat n_o}. 
\eea
The strong coupling ($x\gg1, m/g=$fixed) vacuum of the LSH Hamiltonian is given by:
\bea
n_l(r)&=& 0~~\forall r \nonumber \\
n_i(r)&=&1~,~ n_o(r)~=~1 ~~\mbox{for $r$ odd}\label{SCV_LSH} \\
n_i(r)&=&0~,~ n_o(r)~=~0 ~~\mbox{for ~$r$ even}\nonumber
\eea
which satisfy AGL on all the links. 
The spectrum and dynamics of the LSH Hamiltonian is obtained identical to the gauge invariant dynamics of the Kogut susskind Hamiltonian with the same value of bosonic cut-off using exact diagonalization \cite{Davoudi:2020yln}. The tensor network calculations reported in this work are based on the LSH framework and uses a bosonic cut-off $j_{max}=5/2$, using  the tensor network toolbox \cite{Mathew:2025fim}. 

\subsection*{Approximations for weak coupling regime}
We focus on open boundary condition, and allow an incoming flux $l_i$ to enter the lattice. Towards the weak coupling domain, the interacting vacuum is not dominated by a zero-flux state, SCV, rather it is expected to contain all possible flux states as the electric energy contribution is significantly low, compared to the intereaction energy. This allows us to choose $l_i>>0$, without any loss of generality \cite{Dasgupta:2020itb}. Using AGL on each link, the value of $n
_l$ at each site can be fully determined as
\bea\label{fixnl}
n_l(r)&=&l_i-n_i(r)\left(1-n_o(r)\right)+ \sum_{r'=0}^{r-1}\left(n_o(r')-n_i(r')\right) \nonumber\\
&\approx & l_i-n_i(r)\left(1-n_o(r)\right) ~~\mbox{for, }l_i>>0
.\nonumber
\eea
Thus, any physical state in the LSH formalism in one spatial dimension is completely determined by $(n_i,n_o)$ quantum numbers at each site, and the interaction can be approximated to be local. This approximation is valid \cite{Dasgupta:2020itb} if we focus on the weak coupling regime $x>>1$, and the off-diagonals contribute more to the dynamics. Choosing the global charge sector same as thar of the SCV i.e. with no net Baryon number and no net lattice flux $(\mathcal B,q)=(0,0)$ (see the footnote in main text), the minimum energy strong coupling configuration of the string-ends still remain the same as the SCV. 

We further make an approximation of $$\frac{n_l}{n_l+1 } \rightarrow 1 ~~\& ~~\frac{n_l+1}{n_l+2 }\rightarrow 1$$ for all the prefactors appearing in $H_I^{\mathrm{LSH}}$ considering the  $l_i>>1$ scenario.  

Within this approximation, the total electric part of the LSH Hamiltonian Hamiltonian is given by:
\bea
H_E^{(\mathrm{approx})}=\frac{g^2a}{2}\left[ N h^0_E + \sum_{\{r'\}}\left( \frac{l_i}{2}+\frac{3}{4} \right)\right] 
\eea
where, $\{r'\}$ denotes the sites with fermionic configuration $n_i(r')=0,n_o(r')=1$ and $h^0_E$, corresponds to a global phase due to the background flux. 

\textbf{Mass Hamiltonian:} The mass term (\ref{HMLSH}), being independent of gauge field configuration remain the same in the weak coupling approximation.
\bea
H_M^{(\mathrm{approx})}=\sum_r (-1)^j(\hat n_i(r)+\hat n_o(r))
\eea

\textbf{Interaction Hamiltonian:} The matter-gauge field interaction term is the most complicated within the LSH framework, as detailed in (\ref{HILSH}). In the strong coupling limit of the theory, this particular term gives small contribution to the Hamiltonian. However, in the weak coupling regime of interest, this term becomes significant. 
The approximation scheme that we follow casts the interaction Hamiltonian given in (\ref{HILSH}) as,
\bea
H_I^{(\mathrm{approx})}&=&\sum_r  \Big[ \chi_o^{+}(r)\chi_o^{-}(r+1)
+ \chi_o^{+}(r+1)\chi_o^{-}(r) \nonumber \\ &&
+\chi_i^{+}(r)\chi_i^{-}(r+1)
+\chi_i^{+}(r+1)\chi_i^{-}(r)\Big]
\label{eq: Hi}
\eea

The purpose of the present approximation scheme is to bring the interaction Hamiltonian into simple form, yet describing matter gauge dynamics in the weak coupling regime reliably.




In the next section, we present the quantum algorithm construction, starting with the mapping of LSH quantum numbers to the hardware's qubits and mapping the time evolution unitary for the weak coupling approximated Hamiltonian to a quantum circuit.

\section*{Algorithm for QPU}

In this work we study the time evolution of the system for a scaled time $\tau$ for the scaled Hamiltonian $W$ cause by the unitary operator
\bea
U(t)&=& \exp{-iHt}= \exp{-iW\tau} \\
\mbox{where, } &&W=2xa^3H ~~\Rightarrow~~  \tau = \frac{t}{2xa^3}
\eea

For Trotterized time evolution, the total time $\tau=N_\tau\delta \tau$, where $N_\tau$ denotes the number of Trotter steps. If the value of the coupling changes, i.e. $x\rightarrow x'$, the scaled time \begin{eqnarray}
    \tau\rightarrow \tau'=\frac{x}{x'}\tau ~~
    ~~\Rightarrow ~~ N'_\tau\delta \tau'\, =\, \frac{x}{x'} N_\tau\delta \tau. \label{eq:time}
\end{eqnarray}
If a Trotterized time evolution is performed without changing the size of the Trotter steps (i.e. $\delta \tau'=\delta \tau$), the number of Trotter steps required for evolving the system for a fixed duration of physical time is given by
\begin{eqnarray}
    N'_\tau=\frac{x}{x'}N_\tau. \label{eq:NN'}
\end{eqnarray}

We employ a first-order-Trotterized Schr\"{o}dinger time evolution operator for unitaries constructed for $$xH_I^{\mathrm{approx}}~~,~~ H_E^{\mathrm{approx}}~~\&~~ \mu H_M^{\mathrm{approx}}$$ in that order, over a very short time $\delta_\tau$ for each Trotter step. Evolving the system through $N_{\tau}$ steps span a total time of $\tau=N_{\tau}\delta_\tau$. We choose $\delta_\tau$ to be very short to minimize the Trotterization error and keep the same fixed to 0.0015 throught the work.

\subsection*{Mapping string-ends to qubits}

An open-boundary lattice of $N$ sites is simulated on $2N$ qubits. The numbers $n_i(r)$ and $n_o(r)$ are directly mapped to individual qubits, whereas the $n_l(r)$ numbers can be recovered using (\ref{fixnl}). To make the qubits effectively fermions, the Jordan-Wigner transformation maps the ladder operators:

\begin{equation}\label{ladders}
\begin{aligned}
\hat\chi_{i}^-(r)&=\sigma^+(r)\prod_{r'<r}(-\sigma^z(r'))\\
\hat\chi_{o}^-(r)&=\sigma^+(r+L)\prod_{r'<r+N}(-\sigma^z(r')),
\end{aligned}
\end{equation}
where the qubit ladder operators are defined as
\begin{equation}
\begin{aligned}
\sigma^+=\begin{pmatrix}
0 & 1 \\
0 & 0 
\end{pmatrix}\implies&\sigma^+\ket{0}=\sigma^+\begin{pmatrix}
1 \\
0 
\end{pmatrix}=0,\\
&\sigma^+\ket{1}=\sigma^+\begin{pmatrix}
0 \\
1 
\end{pmatrix}=\begin{pmatrix}
1 \\
0 
\end{pmatrix}.
\end{aligned}
\end{equation}

The Hermitian conjugates are $\hat\chi_{i}^+(r)=(\hat\chi_{i}^-(r))^\dagger$ and $\hat\chi_{o}^+(r)=(\hat\chi_{o}^-(r))^\dagger$, with $\sigma^-=(\sigma^+)^\dagger$.

The implied ordering of the variables where the first $N$ represent the $n_i$ quantum numbers, and the last $N$ the $n_o$ numbers, is relevant only within the mathematical construction, providing, for example, $\hat\chi^+\hat\chi^-$ products and the indices in matrix calculations. When considering the actual mapping to the hardware's qubits, a different ordering can be used. The qubit layout we choose is a zigzag (alternating) sequence of $i$ and $o$ quantum numbers:

\begin{equation}
\ket{q_{2r}} \equiv \ket{n_{i}(r)},\quad\ket{q_{2r+1}} \equiv \ket{n_{o}(r)}.
\end{equation}

This minimizes the number of SWAP layers throughout the circuit. Each $n_i(r)$ number needs to interact with both of its $i$ neighbours, $n_i(r-1)$ and $n_i(r+1)$, and similarly for $o$ numbers, while the electric Hamiltonian term acts on the $i-o$ pairs. All the SWAP gates required to enable this interaction can be arranged in two layers only for each trotter step (see Fig.~\ref{fig:circ}), making this a highly efficient encoding whose depth depends only on the number of trotter steps, and not on the lattice site. Note that, in the initial Trotter step, the first layer of gates would have been the SWAPs to get the first interaction pairs adjacent, so the actual qubit initialization has an ordering that absorbs these SWAPs: $n_i(0)$, $n_i(1)$, $n_o(0)$, $n_o(1)$, $n_i(2)$, $n_i(3)$, $n_o(2)$, $n_o(3)$,..., while the layout at the end of each Trotter step follows the zigzag ordering above.

\subsection*{Building the Quantum Circuit }

A single Trotter step applies the unitary
\bea
U(\delta_\tau,c,\tilde m,\theta)&=&e^{-i\delta_\tau H'} \\
&\;\stackrel{\text{small }\delta_\tau}{\approx}&e^{-i\tilde m H_M}e^{-i\delta_\tau H_E}e^{-icH_I}\nonumber 
\eea
where $c=\delta_\tau x$, $\tilde m=\delta_\tau\mu$ 
are parameters appearing in the corresponding unitaries due to the interaction, and  mass parts of the Hamiltonian. 

The interaction part given in (\ref{eq: Hi}) translates into its qubit version as
\begin{equation}
\begin{aligned}
H_I^{\mathrm{approx}}
&=-\sum_{\substack{r=0\\r\neq N-1}}^{2N-2}\left[\sigma^-(r)\sigma^+(r+1)+\sigma^+(r)\sigma^-(r+1)\right]
\end{aligned}
\label{eq:Hop}
\end{equation}
The corresponding unitary takes the form
\begin{equation}
\begin{aligned}
&U_I=e^{-i\delta_\tau xH_I}\\
&\stackrel{\text{small }\delta_\tau}{\approx}\prod_{r\neq N-1}exp\left[ic\left(\sigma^-(r)\sigma^+(r+1)+\sigma^+(r)\sigma^-(r+1)\right)\right]\\
&\;\;\;\;=\prod_{r\neq N-1}exp\left[ic\begin{pmatrix}
0 & 0 & 0 & 0 \\
0 & 0 & 1 & 0 \\
0 & 1 & 0 & 0 \\
0 & 0 & 0 & 0
\end{pmatrix}_{r\otimes r+1}\right]\\
&\;\;\;\;=\prod_{r\neq N-1}\begin{pmatrix}
1 & 0 & 0 & 0 \\
0 & \cos c & i\sin c & 0 \\
0 & i\sin c & \cos c & 0 \\
0 & 0 & 0 & 1
\end{pmatrix}_{r\otimes(r+1)}
\end{aligned}
\end{equation}
Here $c=\delta_\tau x$ is a real constant. Since each non-boundary-site qubit appears in two of the factors, corresponding to hopping in two directions, the algorithm is split into two layers with SWAP gates in between. Using matrix multiplication, the two-qubit hopping term in (\ref{eq:Hop}) is uncovered in terms of standard single- and two-qubit gates:
\begin{equation}\label{HIfinalform}
\begin{aligned}
U_{I,r\otimes (r+1)}=&CNOT_{(r+1)\otimes r}(1\otimes H)\;CNOT_{r\otimes(r+1)}\\
&(1\otimes R_z(c))CNOT_{r\otimes(r+1)}(1\otimes R_z(-c))\\
&(1\otimes H)\;CNOT_{(r+1)\otimes r}
\end{aligned}
\end{equation}

After the two layers of this hopping unitary, SWAP gates get the qubits into the zigzag position for the electric and mass terms to be applied.

The electric part is
\begin{equation}
H_E=N h^0_E + \sum_{\{r'\}}\left( \frac{n_l}{2}+\frac{3}{4} \right)
\end{equation}
with $\{r'\}$ denoting the set of sites in the fermion configuration $01$ (corresponding to the second entry in the 2-qubit tensor product). The first term gives a global phase to the evolution of qubits, so we ignore it. This is also an approximation because the value of the average flux $h_E^0$ differs between basis states, but we approximate it to always be equal to the boundary flux. However, the phase correction in the $01$ case at every site and its entangling properties can be seen to increase the validity of the simulation. This second term in $H_E$ is translated into a circuit applied on each site's qubit pair, which gives a certain phase in the required $01$ case, and a different phase in the other three cases (same for all three):
\begin{equation}
\begin{aligned}
U_{E,x\otimes (r+1)}(\theta)=&(r\otimes R_z(\theta/2))CNOT_{x\otimes(r+1)}(1\otimes R_z(-\theta/2))\\
&CNOT_{x\otimes(r+1)}(R_z(\theta/2)\otimes 1)(r\otimes 1)\\
=&\begin{pmatrix}
e^{-i\theta/4} & 0 & 0 & 0 \\
0 & e^{i3\theta/4} & 0 & 0 \\
0 & 0 & e^{-i\theta/4} & 0 \\
0 & 0 & 0 & e^{-i\theta/4}
\end{pmatrix}_{x\otimes(r+1)}
\end{aligned}
\end{equation}
which is a global phase plus an $e^{i\theta}$ phase in the desired case. The $x$ and $x+1$ here correspond to the $n_i,\,n_o$ pair of any single lattice site.

$\theta$ can be identified from the desired phase
\begin{equation}
e^{-i\delta_\tau\left( \frac{n_l}{2}+\frac{3}{4} \right)}=e^{i\theta}\;\implies\;\theta=-\delta_\tau\left( \frac{n_l}{2}+\frac{3}{4} \right)
\end{equation}

The mass term $H_M$ adds a phase to each qubit:
\begin{equation}
\mu H_M=\mu\sum_r(-1)^r(n_i(r)+n_o(r))
\end{equation}
\begin{equation}
\begin{aligned}
e^{-i\delta_\tau\mu H_M}=&\prod_re^{i(-1)^{r+1}\tilde mn_i(r)}e^{i(-1)^{r+1}\tilde mn_o(r)}\\
=&\prod_{r\,even}R_z(-\tilde m)_{2r}R_z(-\tilde m)_{2r+1}\\
&\prod_{r\,odd}R_z(\tilde m)_{2r}R_z(\tilde m)_{2r+1}
\end{aligned}
\end{equation}
where $\tilde m=\delta_\tau\mu$. The $R_z$ gates are applied at the end of each Trotter step.


\section*{Proof of Concept}
\begin{figure}[h]
\centering
\includegraphics[width=0.45\textwidth]{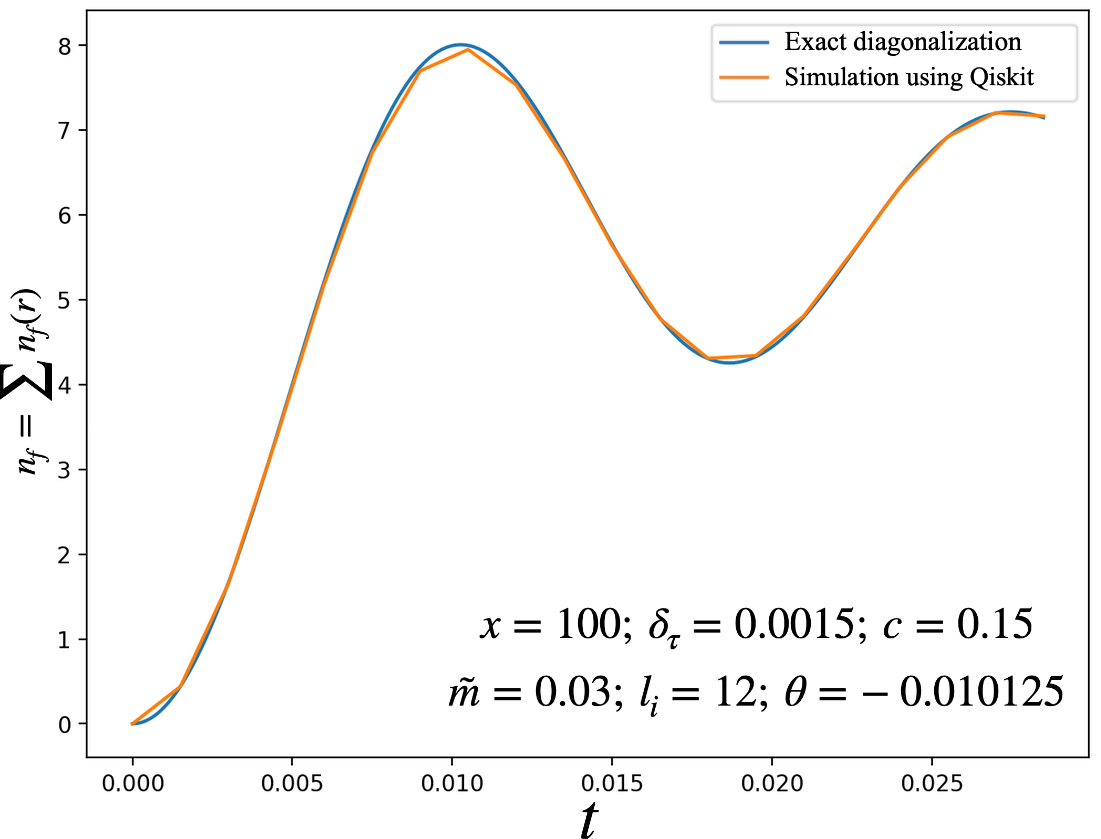}
\caption{\small\label{fig:frog}\textbf{Proof of concept.} Particle number for a lattice of 6 sites as calculated from qubit expectation values at each Trotter step using Qiskit simulator, compared with the exact diagonalization result for the full LSH Hamiltonian, which is free from any Trotterization error and approximation error. Parameters in the quantum circuit are chosen to reproduce the intended regime of the theory with $x=100$ and $m/g=1$.  The initial state is chosen to be the strong coupling vacuum, which is a computational basis state.}
\end{figure}

The algorithm is simulated classically for small numbers of qubits. For a 6-site lattice (12 qubits), ideal simulation of the quantum circuit is compared to exact diagonalization resuts by plotting the total particle number over time for both cases. Exact diagonalization takes the full LSH Hamiltonian as given in (\ref{HELSH}), (\ref{HMLSH}) and (\ref{HILSH}). The calculation of dynamics is also exact via simple matrix multiplication, and free from any Trotterization approximation. Comparison of the dynamics for the original theory and the same obtained via ideal simulation of the quantum circuit is presented in Fig. \ref{fig:frog}. 

\section*{Scaling up the quantum simulation}
\begin{figure*}
    \centering
    \includegraphics[width=1\linewidth]{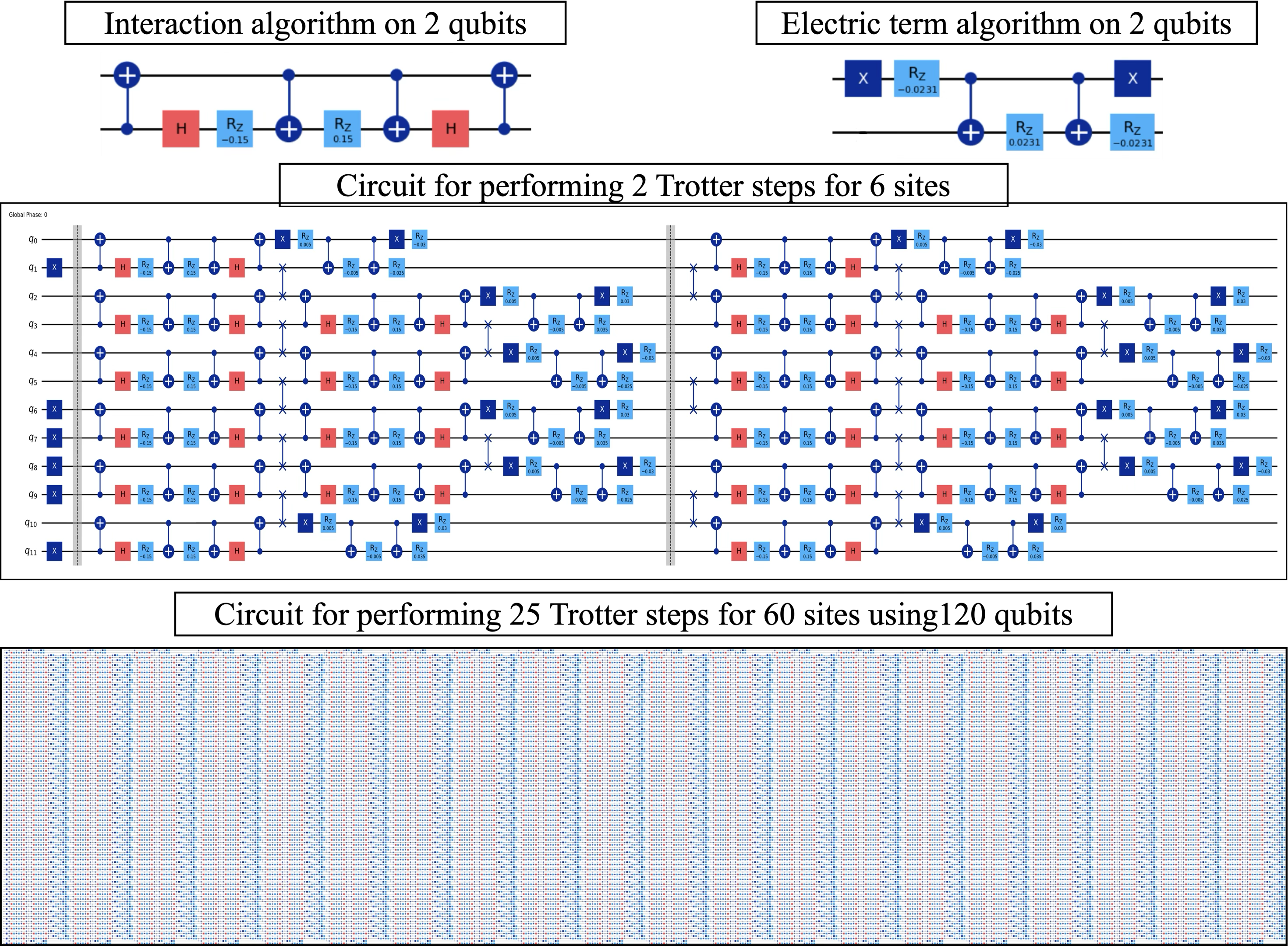}
    \caption{\small\textbf{The Quantum Circuit:} implements unitaries constructed for each term of the Hamiltonian. The mass Hamiltonian is a single qubit rotation, while the building block of electric term and interaction term of the Hamiltonian are 2-qubit operations. Use of a number of swap gates allows simultaneous application of the unitaries for both in a single Trotter step. The middle panel presents two Trotter steps of the full algorithm over 6 lattice sites, initialized with a pair of hadron-antihadron at the center. The bottom panel presents the quantum circuit employed in this study upto 25 Trotter steps of the full algorithm over 60 lattice sites for $x=100$, initialized with a meson at the center. The full circuit consists of $17660$ 2-qubit gates and over $90000$ single-qubit gates. }
    \label{fig:circ}
\end{figure*}

With the benchmarking of the parameters as presented in Fig. \ref{fig:frog}, we proceed towards large scale implementations of the quantum circuit (Fig.~\ref{fig:circ}) using {\it IBM BOSTON} (156 qubit Heron r3 processor).

\subsection*{Experimental measurement from QPU}
We implement trotterized time evolution with a circuit depth that is independent of system size. Each Trotter step comprises two SWAP layers to mediate nearest‑neighbor interactions; for steps $t_{trot} \geq 2$ we prepend a single additional SWAP layer to account for the initial zigzag configuration of qubits. Because the circuit layout is isomorphic to the device topology, qubit mapping requires no extra SWAPs. The Qiskit transpiler is used solely to (i) select a low‑noise linear chain of physical qubits and (ii) decompose the circuit to the native gate set of the device.

At each Trotter step, we estimate $\langle \sigma_z \rangle$ for every qubit and infer the excited‑state occupancy via $P(1) = \frac{1-\langle \sigma_z \rangle}{2}$. Expectation‑value estimation enables error‑mitigation workflows, and keeps the path open for future application of state-of-the-art error mitigation; in this work we apply measurement‑error mitigation to correct readout bias. Following these strategies, we simulate a $60$‑site staggered SU(2) lattice using 120 qubits (two qubits per site). The two‑qubit depth for $t$ Trotter steps is $13t-1$, where the $-1$ accounts for a bypassed SWAP layer enabled by the zigzag qubit initialization. Thus, at $t=25$, the circuit has two‑qubit depth 324 and comprises 17,660 two‑qubit gates with a total gate count of 90,955, the largest reported till date. 

No exact classical method is available at this scale. For validation, we compare the hardware results against approximate tensor‑network simulations and approximate Pauli‑propagation method.

\subsection*{Benchmarking via TN}
In this section, we briefly outline the Matrix Product State (MPS) ansatz used to benchmark the QPU results. Classical Tensor Network (TN) methods have emerged as powerful tools to study properties of low-dimensional quantum systems \cite{Schollwoeck:2010uqf,Banuls:2022vxp}. Of late, they have been used to explore static and dynamic properties of lattice gauge theories, see Ref.\cite{Banuls:2019rao} for a comprehensive review on this topic. In this context, we have developed a tensor network ansatz for the Loop-String-Hadron formulation \cite{Mathew:2025fim}, whereby one can calculate static and dynamic properties of this theory. We will skip the details of the implementations and briefly explain the essentials to establish the workflow.
The ansatz is defined for the full LSH Hamiltonian given in Eq.~\ref{eq:HLSH_full}. Accordingly, all subsequent definitions of the Hilbert space and the associated operators are formulated with respect to the full theory.
The full Hilbert space at each site is characterized by a set of three quantum numbers $\ket{n_l,n_i,n_o}$ and the MPS ansatz is directly endowed with this structure. Formally, one can write it down as follows:
\begin{equation}
    \ket{\Psi[A]} = \sum_{p_1,\ldots,p_N}A^{a_1}_{p_1}A^{a_1,a_2}_{p_2}\ldots A^{a_{N-1}}_{p_N}\ket{p_1,p_2,\ldots,p_N}.
\end{equation}
Here, $p_r$ corresponds to the physical states at each lattice site $r$. The notation $\ket{p_1,p_2,\ldots,p_N}\equiv\otimes_{r=1}^N\ket{p_r}$, where $\ket{p}_r \equiv\ket{n_l,n_i,n_o}_r$. The variational degrees of freedom are contained in the $A$-matrices, which is defined at each sites and populated by complex entries. We further impose the two global symmetries of the LSH Hamiltonian, namely $\mathcal{B}$ and $q$ into the local tensor structure to yield a blocksparse representation.
We rely on the ITensors.jl library \cite{itensor,itensor-r0.3} and its inbuilt functions to construct the MPS/MPO functions. For time-evolving the initial state, we use the 2-site time-dependent varioanal principle (TDVP) algorithm \cite{Paeckel:2019yjf} defined in the library. Due to computational overhead with the 2-site algorithm, we have opted to keep the maximum bond-dimension to 200. This will result in errors accumulating as the time evolution progresses, and is reflected in the entanglement entropy plot in the Supplementary figures. Tabulated values this error is presented in the Supplementary information. The simulation is set up for each value of $x \in \{50,100,200 \}$ by first initializing two product state MPSes, one for the strong-coupling vacuum state and the other for the string-configurations. Both of these states correspond to the same symmetry sector of the LSH Hamiltonian and the algorithm trivially conserves these quantum numbers by construction. The relevant observables are computed for both these time evolved states and subtract to isolate the evolution of the initialized meson.

\subsection*{Benchmarking via PP}
We benchmark quantum processing unit (QPU) observables using the Pauli Propagation Method (PP) \cite{Rudolph:2025gyq}, a Heisenberg‑picture simulator that back‑propagates the measured observable $O$ through the circuit partitioned into logical layers. Under Clifford evolution, Pauli operators map to Pauli operators; consequently, purely Clifford layers preserve the operator’s Pauli support and do not increase the number of qubit-wise non‑commuting terms. By contrast, a non‑Clifford layer whose generator anticommutes with a Pauli component of $O$ can induce branching: in the worst case, the number of Pauli terms doubles across such a layer, leading to exponential growth in the term count with circuit depth. To bound the computational cost, one may truncate by discarding Pauli terms whose coefficients fall below a prescribed threshold during back‑propagation, trading accuracy for tractability in a controlled manner.

In this work, we target non‑truncated PP, i.e., back‑propagation proceeds without discarding terms, rendering the result effectively exact up to floating‑point tolerances. Since there is no immediate way to identify the number of Pauli terms required to backpropagate the circuit completely and exactly, to set a safe compute budget, we empirically upper‑bounded the Pauli‑term cardinality at the circuit input as follows: we tracked (i) the number of Pauli terms and (ii) the remaining layer depth over an initial segment of the back‑propagation, then extrapolated to depth zero. The inferred terminal term counts were 9,400, 12,500, and 66,000 for $x = 50, 100$ and $200$, respectively. Within these budgets, back‑propagation is \emph{expected} to be completed without, or minor, truncation. Accordingly, the PP estimates serve as reference values for the corresponding QPU observables at these settings and scales.


We also implement the PP algorithm on GPUs using NVIDIA's \texttt{cuQuantum} library, specifically  the
\texttt{cuPauliProp} API together with CUDA for memory management and
parallel execution \cite{10313722, NVIDIAcuPauliPropDocs}. 
In PP method, the observables are represented as Pauli expansions
\(
O = \sum_k c_k P_k,
\)
where \(P_k\) are packed Pauli strings and \(c_k\) are real coefficients.
Starting from an initial local observable such as \(Z_q\) acting on qubit
\(q\), the PP algorithm repeatedly conjugates the operator with the circuit
gates, which generates a growing linear combination of Pauli strings
represented on the GPU using \texttt{cupaulipropPauliExpansion} data
structures. Since the number of Pauli terms increases rapidly with circuit depth,
we employ a coefficient-based truncation strategy. After each operator
application, Pauli terms whose coefficients satisfy
\(
|c_k| < \epsilon
\)
are discarded, where we choose the cutoff
\(
\epsilon = 10^{-5}.
\)
To maintain a compact representation of the Pauli expansion,
the GPU performs a parallel radix sort to group identical Pauli strings,
followed by a parallel reduction that merges duplicates and sums their
coefficients. During this compaction step, terms with coefficients below
the cutoff are removed. This procedure trades a controlled amount of
numerical precision for substantial reductions in memory usage and runtime.
The truncation is implemented using the
\texttt{CUPAULIPROP\_TRUNCATION\_STRATEGY\_COEFFICIENT\_BASED}
option provided by the \texttt{cuPauliProp} library. All Pauli expansions, coefficient arrays,
and intermediate buffers remain resident in GPU memory to avoid
host--device data transfers, while dedicated scratch workspace buffers are
allocated for the operator-application kernels. After the observable has
been propagated through the circuit, its expectation value is evaluated in
the computational zero state, (representing the strong coupling vacuum)
\(
\langle O \rangle =
\langle 0^{\otimes n} | O(t) | 0^{\otimes n} \rangle,
\)
where only Pauli strings composed entirely of $Z$ and identity operators
contribute to the trace. This trace is computed on the GPU using the
available routine.
The complete simulation workflow therefore consists of generating a
Trotterized quantum circuit, initializing the observable
\(O = Z_q\) for each qubit \(q\), constructing a single-term Pauli
expansion, propagating the operator through the circuit in reverse order
with truncation applied after each gate, computing the expectation value
\(\langle Z_q(t) \rangle\), and repeating the procedure for all qubits and
Trotter steps. This GPU-based implementation enables efficient evaluation
of local observables in large quantum circuits while controlling the
exponential growth of the Pauli expansion through coefficient truncation.
\subsection*{Estimating Errors}
In this work, we report error bar for the dynamical observable particle density $n_f(t)$ defined as the sum of particle-antiparticle number at each lattice sites $n_f(r,t)$. 
With the quantum experiment using a QPU, and the three approximate classical benchmarks, we obtain four independent estimates for the observable, (without any prior knowledge of the true value)
$$
n_f^{(m)}(t),\, m\in\{\mathrm{TN},\mathrm{PP-CPU},\mathrm{PP-GPU},\mathrm{QPU}\}.$$
The median of these distributions is chosen as the refernce, along with the standard deviation of these 4 data sets at each time slices, 
\begin{eqnarray}
\sigma_{\mathrm{std}}(t)&=&\sqrt{\frac{1}{3}\sum_m\Big(n_f^{(m)}(t)-\overline{n_f}(t)\Big)^2},\\
\mbox{with } ~~
\overline{n_f}(t)&=&\frac{1}{4}\sum_m n_f^{(m)}(t),\nonumber \\ 
\end{eqnarray}
 denoting an intrinsic method-to-method uncertainty.
We further note that, the observable $n_f(r,t)$ is directly related to measuring $n_i+n_o$ excitations at each site while the same quantity contributes to a conserved global charge $\mathcal B+N=Q=\sum_{r} n_i(r)+n_o(r)=60$. We monitor global-symmetry conservation through the total charge
\begin{equation}
 \Delta Q^{(m)}(t)=Q^{(m)}(t)-60,
\end{equation}
and convert this constraint violation into an additional error scale.  Assuming $Q$ is a sum over $L=60$ sites, we use the conservative per-site scaling $\Delta Q^{(m)}(t)/\sqrt{L}$ and define a method-aggregated constraint term by the RMS
\begin{equation}
\Delta Q_{\mathrm{rms}}(t)=\sqrt{\frac{1}{4}\sum_m\big(\Delta Q^{(m)}(t)\big)^2}.
\end{equation}
Our final reported reference value with shared uncertainty band ($n_f^{\mathrm{ref}}(t)\pm\sigma_{\mathrm{shared}}(t)$) around the reference curve is obtained by quadrature,
\begin{equation}
\sigma_{\mathrm{shared}}(t)=\sqrt{\sigma_{\mathrm{std}}(t)^2+\left(\frac{\Delta Q_{\mathrm{rms}}(t)}{\sqrt{60}}\right)^2},,
\end{equation}
which captures both cross-method variability and global-constraint inconsistency.  

Additionally, for each method we report individual error bars
$$\sigma_m(t)=\sqrt{\sigma_{\mathrm{m,\mbox{pair}}}(t)^2+\left(\frac{|\Delta Q^{(m)}(t)|}{\sqrt{60}}\right)^2},$$ where
$$\sigma_{\mathrm{m,\mbox{pair}}}(t) = |n_f^{(m)}(t)- n_f^{\mathrm{ref}}(t)|.$$
The deviations from global conservation inflate the uncertainty in a controlled and transparent manner.
\subsection*{First excitation energy gap of Hadron}
The internal breathing-mode frequency of the length-one meson was extracted from the site-resolved quark-antiquark occupation data $n_f(r,t)$ by three independent strategies:
\begin{enumerate}
    \item 
 Centering the profile on the initial meson bond and constructing the symmetrized central-shell observable
$$
\rho_m(t)=\frac{n_{f}(30,t)+n_{f}(31,t)}{2},
$$
which measures the time-dependent weight on the central bond of the meson.
\item Focussing on the central lattice region
$$
r \in \mathcal{W}=\{23,24,\dots,38\},
$$
and defining the second-moment observable
\begin{equation}
R^2(t)=
\frac{\sum\limits_{r\in\mathcal{W}} (r-r_0)^2\,|n_f(r,t)|}
{\sum\limits_{r\in\mathcal{W}} |n_f(r,t)|},
\qquad r_0=30.5. \nonumber
\end{equation}
The absolute value is used so that the sign-changing red/blue oscillation in $n_f(r,t)$ does not artificially cancel the spatial weight.\footnote{
A full-lattice definition of $R^2(t)$ was checked separately, but it was found to be unstable because the factor $(r-r_0)^2$ strongly amplifies small far-tail/background contributions. The central-window definition therefore gives the cleanest extraction of the internal oscillation.}
\item Considering simply the oscillation of total number of fermion and anti-fermions $n_f(t)=\sum_{r=1}^{60} n_f(r,t)$. 
\end{enumerate}
Each of the oscillatory functions are plotted in a stretched-exponential damped oscillatory form
$$
\rho_0(t)=C+A\,e^{-b t^{\alpha}}\cos(\omega t+\phi),
$$
and identified \(\omega\) as the breathing-mode frequency, equivalently the estimate of the meson gap \(E_1-E_0\) in units of inverse time step. 

The frequency obtained by the second method is found to be roughly the double of the frequency obtained by the first and third method (mutually consistent) as the second moment is only aware of the absolute value for the observables. For $x=100$, TN, PP-GPU and QPU provides the frequencies to be $1.1765\pm0.0324, 1.1880\pm 0.0263,
1.1714\pm 0.0266$, whereas the one obtained by the third methos is reported in the main text. It is crucial to note that the frequency extracted from QPU and its error bar is of the exact same order as obtained by the classical methods.
 In the broader comparison across methods and various couplings, the extracted frequencies increase approximately linearly with coupling. The final values used in the comparison plots are listed in Table~\ref{tab:breathing_mode_frequencies}.
\begin{table}[h]
\centering
\begin{tabular}{@{}cccc@{}}
\toprule
\textbf{\(x\)} & \textbf{QPU} & \textbf{PP-CPU} & \textbf{TN} \\
\midrule
50  & 0.20036  & 0.28720  & 0.288129 \\
80  & 0.491033 & 0.483248 & 0.484251 \\
100 & 0.59100  & 0.594675 & 0.599000 \\
200 & 1.20634  & 1.21320  & --- \\
\bottomrule
\end{tabular}
\caption{Extracted breathing-mode frequencies $\omega$ for different couplings $x$ and using different hardware/ methods. A dash indicates that no reliable frequency could be extracted.}
\label{tab:breathing_mode_frequencies}
\end{table}

\section*{End Notes}

The foundation of the current experiment lie on LSH encoding, followed by the weak coupling approximation detailed in the Methods. This approximation is essential to come up with an implementable quantum algorithm for the complicated dynamics using restricted number of qubits with restricted connectivity on the \emph{IBM Heron} processors. A very important conclusion from the outcome of classical benchmarking is establishing the validity of the approximation for the coupling regime of interest. The value of observables obtained via the PP (classical GPU simulation of the approximated LSH Hamiltonian) and the same obtained via TN (classical simulation of the full LSH Hamiltonian) agree up to an average standard deviation of $0.01$ over all time slices. 

The average value of the error bar (defined for cross-method validation) associated with the observable $n_f$ for QPU is obtained as $0.0973$, while the same for PP-CPU is $0.0406$. The TN and PP-GPU have relatively smaller error bars, with average magnitudes of $0.0184$ and $0.0128$, respectively. However, given that all four error bars are of the same order for the reported observable, the quantum data without active error mitigation may still be useful for extracting physical information. 

A technical remark on imposing the bosonic cut-off in this computation is worth mentioning here. In the LSH framework, the cut-off corresponds to the maximum amount of gauge flux allowed on each link. The approximation scheme is set up for an arbitrarily large value of the cut-off as we consider a large amount of flux to enter and exit the lattice. The same is equally valid for the quantum algorithm and its implementation using a QPU and a PP. However, the TN algorithm can only work with a finite cut-off; we set the cut-off to $6$ for the TN benchmark. The value of the phase angle $\theta$ in the quantum algorithm is chosen to match this limitation on the cut-off. In principle, this particular quantum algorithm allows computations to be performed with a larger cut-off, and goes beyond the scope of TN calculations.

The quantum processor maintains structural robustness, as reflected in the consistent oscillatory behaviour of the number density over time.  Violation of global symmetries, albeit minimal for PP and QPU both, grows systematically with increasing $x$ for longer time for PP (see supplimentary figures for details), while remaining of the same order for QPU at all values of $x$ and at all time scales. A careful look into the high fidelity results from QPU suggests a lightcone to emerge as an outcome of the differential measurement for all values of x, similar to the same obtained via PP. This is indeed nontrivial given the scale of the circuit it implements (see Fig. \ref{fig:circ}). 

 
These observations suggest the possibility of a robust quantum simulation strategy for notoriously difficult non-Abelian gauge theories and point towards a road map for useful quantum advantage once hardware noise subsides. With the current noisy hardware, some of the physical quantities, such as the frequency of the breathing mode of hadron dynamics can be extracted precisely using QPU, even in a regime where the state-of-the-art classical algorithm (TN \cite{Gupta:2026tcg} and its implementation (with finite budget of bond-dimension or compute time) fail. 



\newpage
{\Huge{\textbf{\noindent Supplementary \\~~Information}}}
\subsection*{Quantum simulation vs. classical simulation}
The details of the runtime for each Trotter step using each method for $x=100$ are tabulated in Table \ref{tab:data_comparison_times_only}.

\begin{table}[h]
    \centering
    \caption{Data Comparison: Execution Times}
    \label{tab:data_comparison_times_only}
    \resizebox{0.35\textwidth}{!}{%
        \begin{tabular}{@{}ccccc@{}}
            \toprule
            \textbf{Trotter} & \textbf{QPU} & \textbf{TN} & \textbf{PP-CPU} & \textbf{PP-GPU} \\ 
            \textbf{step} & \textbf{(sec)} & \textbf{(sec)} & \textbf{(sec)} & \textbf{(sec)} \\ 
            \midrule
            1  & 20 & 170.106  & 23.2512   & 90.2822   \\
            2  & 20 & 200.972  & 114.6801  & 199.2602  \\
            3  & 20 & 290.948  & 232.8814  & 312.8630  \\
            4  & 20 & 423.376  & 352.9601  & 428.8070  \\
            5  & 20 & 584.092  & 477.4471  & 547.5810  \\
            6  & 20 & 786.466  & 592.9340  & 668.0500  \\
            7  & 20 & 1180.006 & 726.0387  & 792.5720  \\
            8  & 20 & 1994.450 & 850.0756  & 919.3930  \\
            9  & 20 & 3223.432 & 981.8030  & 1051.3070 \\
            10 & 20 & 4936.112 & 1111.2079 & 1186.1290 \\
            11 & 20 & 7616.980 & 1246.4771 & 1327.1470 \\
            12 & 20 & 9408.224 & 1355.9682 & 1476.8920 \\
            13 & 20 & 9408.224 & 1517.3462 & 1638.0440 \\
            14 & 20 & 9457.832 & 1606.2973 & 1809.1480 \\
            15 & 20 & 11300.992& 1798.2811 & 1990.1430 \\
            16 & 20 & 10530.030& 1912.3661 & 2173.8900 \\
            17 & 20 & 8936.264 & 2054.6526 & 2357.6000 \\
            18 & 20 & 8905.564 & 2233.0782 & 2543.8400 \\
            19 & 20 & 9148.052 & 2351.2184 & 2734.2200 \\
            20 & 20 & 8949.110 & 2484.3693 & 2930.4400 \\
            \bottomrule
        \end{tabular}%
    }
\end{table}

The estimated error in the tensor network calculation is directly related to the errors building up as the time evolution progresses due to finite bond dimension. This error corresponds to the sum of the Schmidt values that were discarded after each time step as tabulated in Table \ref{tab:mps}.
\begin{table}[h!]
    \centering
    \begin{tabular}{cccc}
        \hline
        Trotter step & \multicolumn{3}{c}{Truncation Error} \\
        & $x=200$ & $x=100$ & $x=50$ \\
        \hline
         1  & $10^{-11}$ & $10^{-11}$ & $10^{-11}$ \\
         2  & $10^{-11}$ & $10^{-11}$ & $10^{-11}$ \\
         3  & $10^{-11}$ & $10^{-11}$ & $10^{-11}$ \\
         4  & $10^{-11}$ & $10^{-11}$ & $10^{-11}$ \\
         5  & $10^{-11}$ & $10^{-11}$ & $10^{-11}$ \\
         6  & $10^{-9}$  & $10^{-11}$ & $10^{-11}$ \\
         7  & $10^{-7}$  & $10^{-11}$ & $10^{-11}$ \\
         8  & $10^{-6}$  & $10^{-11}$ & $10^{-11}$ \\
         9  & $10^{-4}$  & $10^{-11}$ & $10^{-11}$ \\
         10 & $10^{-3}$  & $10^{-11}$ & $10^{-11}$ \\
         11 & $10^{-3}$  & $10^{-10}$ & $10^{-11}$ \\
         12 & $10^{-3}$  & $10^{-9}$ & $10^{-11}$ \\
         13 & $10^{-2}$  & $10^{-8}$ & $10^{-11}$ \\
         14 & $10^{-2}$  & $10^{-7}$ & $10^{-11}$ \\
         15 & $10^{-2}$  & $10^{-6}$ & $10^{-11}$ \\
         16 & $10^{-3}$  & $10^{-5}$ & $10^{-11}$ \\
         17 & $10^{-3}$  & $10^{-5}$ & $10^{-11}$ \\
         18 & $10^{-3}$  & $10^{-4}$ & $10^{-11}$ \\
         19 & $10^{-3}$  & $10^{-4}$ & $10^{-11}$ \\
         20 & $10^{-3}$  & $10^{-4}$ & $10^{-11}$ \\
        \hline
    \end{tabular}
    \caption{The maximum truncation error in the Schmidt values for the time evolved state at each TDVP time step}
    \label{tab:mps}
\end{table}

The values of the individual error bars reported in Fig. 2 of the main text are listed in the  Table \ref{tab:errorbar}.

\begin{table}[t!]
    \centering
    \begin{tabular}{ccccc}
        \hline
        Trotter step & \multicolumn{4}{c}{Magnitude of error bars for} \\
  & TN & PP-CPU & PP-GPU & QPU \\
\hline
1 & 0.000393 & 0.000393 & 0.000394 & 0.037981 \\
2 & 0.001433 & 0.002093 & 0.001433 & 0.078603 \\
3 & 0.007679 & 0.001600 & 0.001600 & 0.004534 \\
4 & 0.014291 & 0.004468 & 0.004468 & 0.052748 \\
5 & 0.022174 & 0.009427 & 0.009427 & 0.077490 \\
6 & 0.030052 & 0.016143 & 0.016143 & 0.157964 \\
7 & 0.036875 & 0.023230 & 0.023230 & 0.170741 \\
8 & 0.041701 & 0.029011 & 0.028255 & 0.028320 \\
9 & 0.007022 & 0.068981 & 0.007037 & 0.016977 \\
10 & 0.007009 & 0.068038 & 0.007017 & 0.170568 \\
11 & 0.005745 & 0.063114 & 0.005752 & 0.191280 \\
12 & 0.003736 & 0.056778 & 0.003827 & 0.174985 \\
13 & 0.002753 & 0.051557 & 0.003012 & 0.161465 \\
14 & 0.003788 & 0.049230 & 0.003976 & 0.013341 \\
15 & 0.034967 & 0.023118 & 0.023093 & 0.049691 \\
16 & 0.042817 & 0.029218 & 0.029168 & 0.130727 \\
17 & 0.049697 & 0.038766 & 0.038798 & 0.090731 \\
18 & 0.042992 & 0.061825 & 0.037303 & 0.038343 \\
19 & 0.000899 & 0.111911 & 0.002302 & 0.081743 \\
20 & 0.011387 & 0.103556 & 0.011421 & 0.239686 \\
\hline
\end{tabular}
\caption{List of the values of the error bars for each method per time slice for computation at $x=100$. }
\label{tab:errorbar}
\end{table}
\vspace{100cm}

\newpage
\newpage 

\begin{figure*}[h!]
    \centering
\includegraphics[width=1\linewidth]{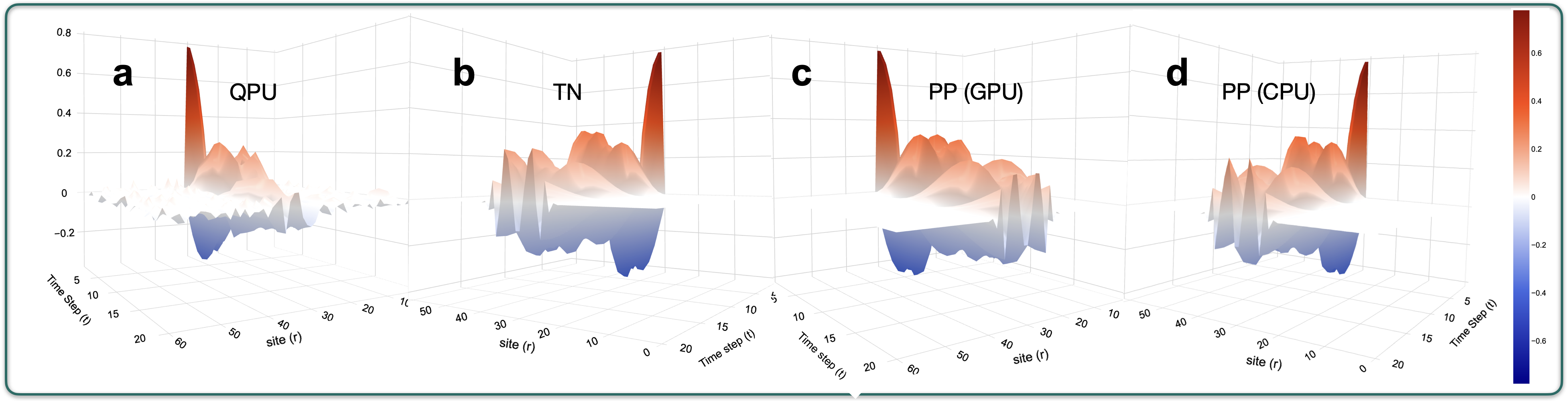}
  \caption{\small\textbf{3-dimensional representation of mesonic profile} Space-time evolution of quark (anti-quark) number $n_f(r,t)$, starting from an initial meson placed at the center of the lattice is obtained on a 60-site lattice.  (a)-(d) represents its 3-D representation, demonstrating internal oscillation and spreading. (a) The noise in QPU is presented as background fluctuation. }
    \label{fig:HGplot}
\end{figure*}

\begin{figure*}
    \centering
    \includegraphics[width=0.9\linewidth]{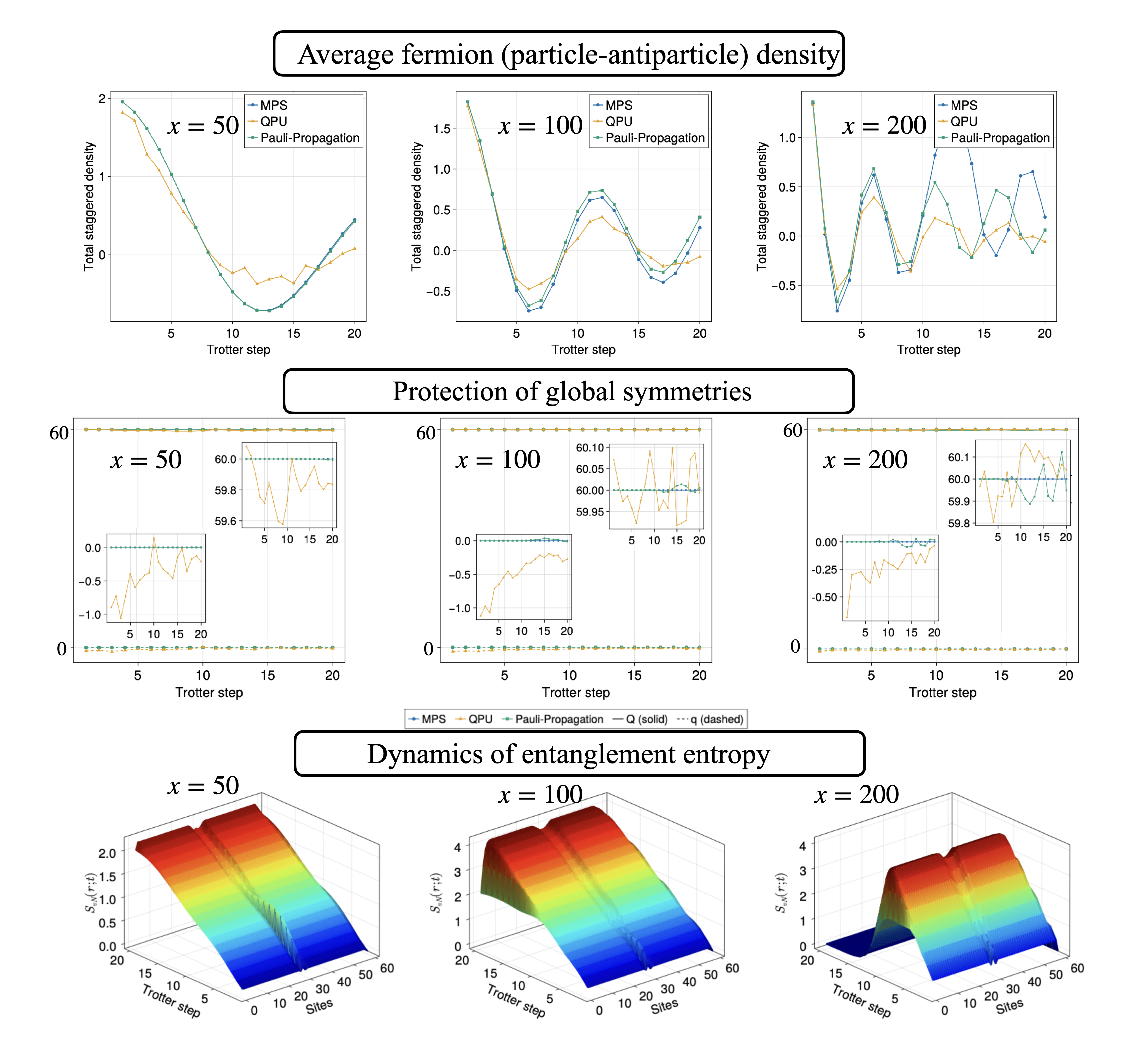}
    \caption{\small{\textbf{Comparing robustness of quantum simulation versus classical simulation towards $x\rightarrow \infty$. }} Top row: The dynamics of average fermion density is plotted with time. For $x=50$, MPS and PP agree exactly, QPU shows deviation but follow the trend. For $x=100$, MPS and PP start to separate out after 5th Trotter steps, QPU deviates but follows the trend.  For $x=200$, post 10th Trotter step, MPS and PP do not follow the same trend, QPU follows PP, with deviation. Middle row: The conservation of global charges are tracked for all simulation. Overall it stays conserved in all the methods. As presented in the insets, QPU shows a small deviation, for all $x$ values. MPS preserves the global constraints by construction \cite{Mathew:2025fim}. PP shows small deviation, and it increases with increasing $x$. Bottom row: The entanglement entropy, obtained via TN calculation, shows linear growth for $x=50$, saturates to a plateau for $x=100$, while the allowed bond dimension fails to handle entanglement growth for $x=200$ case, and in effect the MPS simulation fails to capture physics anymore. The drop in entanglement is an artifact of finite bond dimension, rather than being a physical phenomenon. }
    \label{fig:comp-x}
\end{figure*}

\begin{figure*}[h!]
    \centering
    \includegraphics[width=1\linewidth]{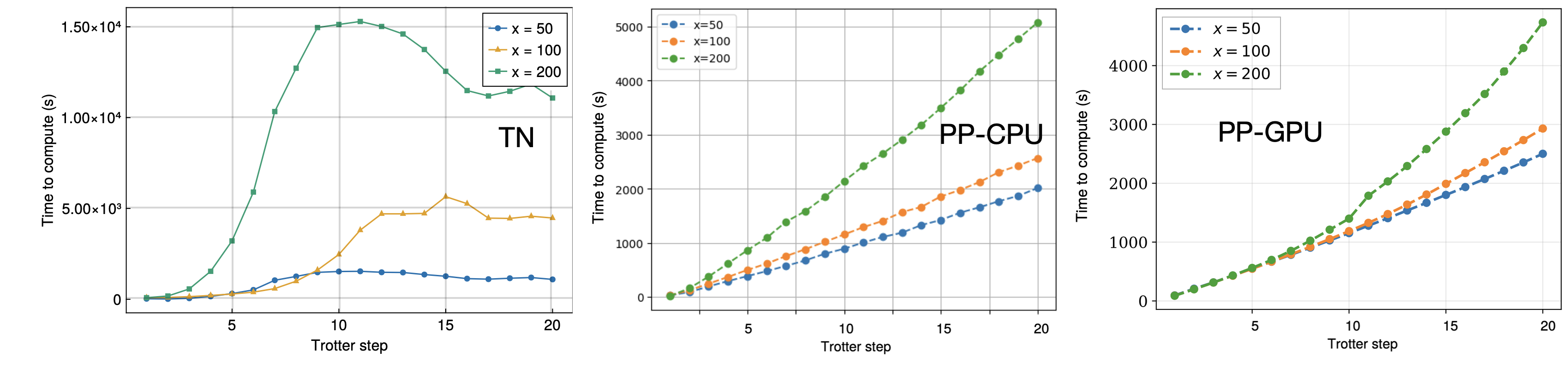}
    \caption{\textbf{Supplementary Figure: Growth of classical computation cost towards the weak coupling limit.} For a fixed system size and the fixed Hamiltonian encoded in MPO or the quantum circuit with a fixed gate depth, the computation time increases as the coupling constant $x$ increases. Clock time for computing each Trotter step of the TN calculation on a Mac Pro tower (3.2GHz 16‐core 4.4GHz Intel Xeon W processor, 96GB) is presented in the left panel. Compute time for Pauli propagation using HP Z2 Tower G9 Workstation id presented in the middle panel. PP on NVIDIA RTX 5000 Ada GPU(AD102GL) in presented in the right panel. }
    \label{fig:placeholder}
\end{figure*}

\newpage

\begin{figure*}[h!]
    \centering
    \includegraphics[width=1\linewidth]{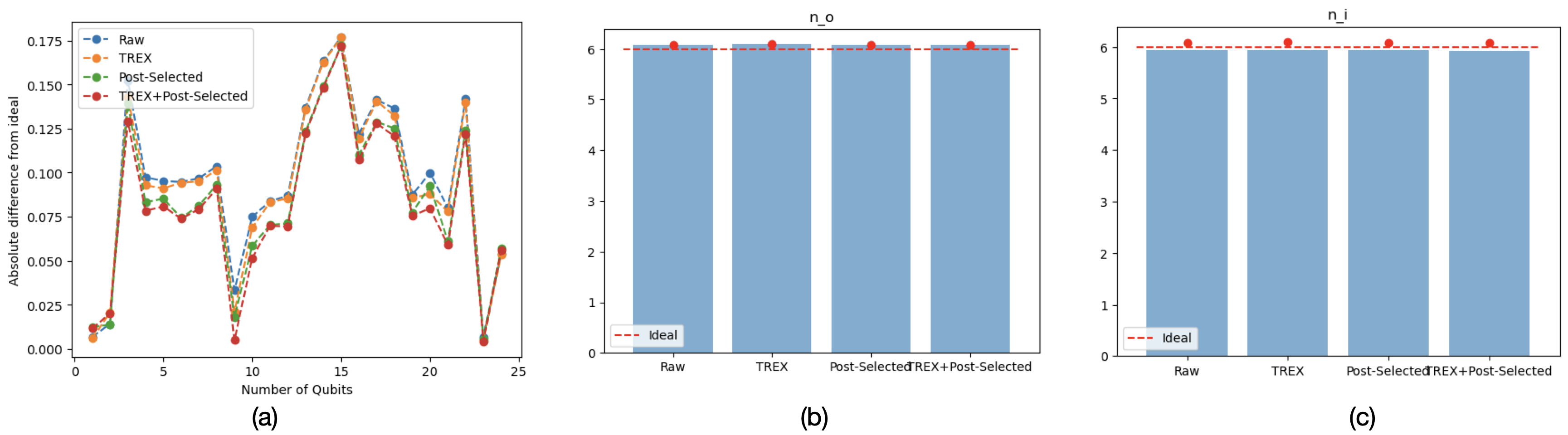}
    \caption{\textbf{Supplementary Figure: POC for errror mitigation strategy employed.} The output from the QPU is presented here with only measurement-error mitigation (TREX). This is decided based on comparing the outcome of simulated noise with the ideal result for a system of $24$ qubits.  (a) Absolute difference from the ideal result for raw, measurement error mitigation (TREX), post-selection and using both are compared. It appears that TREX alone is able to provide good results. (b), (c) Globally conserved charges stay robust with all the methods. }
    \label{fig:mitigation}
\end{figure*}

\begin{figure*}
    \centering
    \includegraphics[width=1\linewidth]{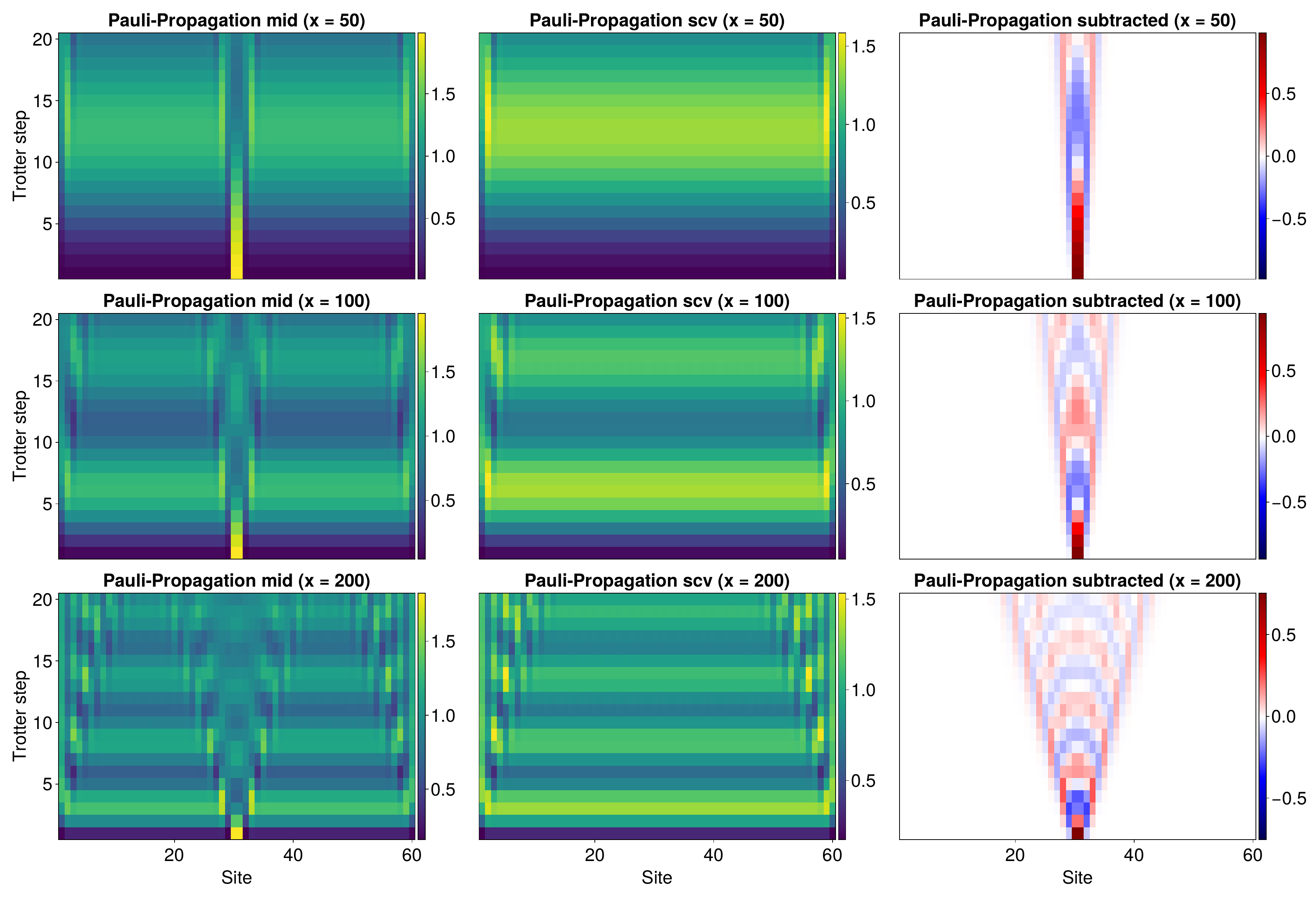}
    \caption{\textbf{Supplementary Figure: Demonstrating differential measurement protocol demonstrated for PP.} One major reason to obtain reasonably good result only with measurement error mitigation is due to the differential measurement protocol being employed in this work. This protocol  cancels some of the systematic error or biases present in the hardware result. In addition, it completely diminishes any boundary effect present in the calculation. As the hadron spreads, performing calculation with a large lattice becomes more and more crucial to avoid any boundary effect. 
The leftmost column represents propagation of hadron, placed at the center of the finite size lattice for theories with coupling constant $x=50,\,100,\,200$. The middle column represents dynamical evolution of the SCV for all three cases. The rightmost column represents the difference between the first two column, which is presented as the result in this current work. It is evident that boundary effects dominate for larger value of x, and hence working with a large lattice is crucial to approach $x\rightarrow \infty$ limit. While this demonstration is with the simulation of noiseless circuit, for the experimental measurement on noisy hardware, some systematic error, mostly if any bias is present gets cancelled and in effect high fidelity results are obtained withut active error mitigation. }
    \label{fig:diff}
\end{figure*}

\end{document}